\documentclass{arima-en}
\pdfoutput=1 

\usepackage[T1]{fontenc}
\usepackage[utf8]{inputenc}
\usepackage{lmodern}
\usepackage{amsfonts}
\usepackage{MnSymbol}
\usepackage{tabularx}  
\usepackage{varwidth}  
\usepackage{adjustbox}
\usepackage{enumitem}
\usepackage[mathscr]{eucal}
\let\savemathscr=\mathscr
\usepackage{mathrsfs}

\let\mathscr=\savemathscr
\let\savemathscr=\relax
\usepackage{ifthen}
\usepackage{graphicx}
\usepackage{wrapfig}
\usepackage{fancyhdr}
\usepackage{hyperref}
\usepackage[babel]{csquotes}

\title{An experimental evaluation of choices of SSA forecasting
  parameters}
\author[*1]{Teodor KNAPIK}
\author[2]{Adolphe RATIARISON}
\author[2]{Hasina RAZAFINDRALAMBO}
\affil[1]{ISEA, Universit\'e de la Nouvelle Cal\'edonie, France} 
\affil[2]{DyACO, Universit\'e d'Antananarivo, Madagascar} 
\corrauthor{knapik@unc.nc}

\doi{10.46298/arima.9641}{}
\review{June 1, 2022}{February 29, 2024}
\publication{40}{2024}
\editors{Mathieu Roche, Nabil Gmati, Clémentin Tayou Djamegni}
\usepackage{lastpage}
\usepackage[
  style=numeric-comp,
  datamodel=software, 
  abbreviate=false,
  natbib=true,
  sorting=ynt,
  backend=biber,
  bibencoding=utf8,
  giveninits=true,
  url=false,
  doi=false,
  defernumbers,
  maxcitenames=10,
  defernumbers=false,
  maxbibnames=100]{biblatex}
\usepackage{software-biblatex}
\newcommand{\doiorurl}{%
 \iffieldundef{doi}
    {\iffieldundef{url}
       {}
       {\strfield{url}}}
    {http://dx.doi.org/\strfield{doi}}%
}
\newcommand{\myhref}[1]{%
 \ifboolexpr{%
  test {\ifhyperref}
  and
  not test {\iftoggle{bbx:url}}
   and
   not test {\iftoggle{bbx:doi}}
  }
  {\href{\doiorurl}{#1}}
  {#1}%
}
\DeclareFieldFormat{title}{\myhref{\mkbibemph{#1}}}
\DeclareFieldFormat
  [article,inbook,incollection,inproceedings,patent,thesis,unpublished]
  {title}{\myhref{\mkbibquote{#1\isdot}}}
\newcommand{\bbbn}{{\mathrm{I\!N}}}

\newcommand{\bbbr}{{\mathrm{I\!R}}}

\renewcommand{\setminus}{\smallsetminus}%
\newcommand{\tsx}{\mathbb{X}}
\newcommand{\tsy}{\mathbb{Y}}
\newcommand{\tsz}{\mathbb{Z}}
\newcommand{\trx}{\mathbf{X}}
\newcommand{\utrx}{\underline{\trx}}
\newcommand{\bU}{\mathbf{U}}

\newcommand{\uU}{{\underline{U}}}
\newcommand{\ubU}{{\underline{\bU}}}
\newcommand{\cU}{\mathcal{U}}
\newcommand{\ucU}{{\underline{\cU}}}
\newcommand{\bV}{\mathbf{V}}
\newcommand{\bSg}{\mathbf{\Sigma}}
\newcommand{\ttrx}{\tilde{\trx}}
\newcommand{\oI}{{\overline{I}}}
\newcommand{\ox}{{\overline{x}}}
\newcommand{\ba}{\mathbf{a}}
\newcommand{\be}{\mathbf{e}}

\newcommand{\bF}{\mathbf{F}}
\newcommand{\bxi}{\boldsymbol{\xi}}
\newcommand{\bx}{\mathbf{x}}
\newcommand{\by}{\mathbf{y}}
\newcommand{\bz}{\mathbf{z}}
\newcommand{\bY}{\mathbf{Y}}
\newcommand{\bw}{\mathbf{w}}
\newcommand{\mean}{\mathrm{mean}}
\newcommand{\rnk}{\mathrm{rank}}
\newcommand{\calV}{\mathcal{V}}
\newcommand{\calJ}{\mathcal{J}}
\newcommand{\nbs}{\kern .2em}
\begin{document}
\maketitle

\abstract{ Six time series related to atmospheric phenomena are used
  as inputs for experiments of forecasting with singular spectrum
  analysis (SSA). Existing methods for SSA parameters selection are
  compared throughout their forecasting accuracy relatively to an
  optimal \emph{a posteriori} selection and to a naive forecasting
  methods.  The comparison shows that a widespread practice of
  selecting longer windows leads often to poorer predictions.  It also
  confirms that the choices of the window length and of the grouping
  are essential.  With the mean error of rainfall forecasting below
  $1.5\%$, SSA appears as a viable alternative for horizons beyond two
  weeks.
}

\keywords{time series forecasting, singular spectrum analysis,
  parameter selection}

\section{Introduction}
Many naturally occurring dynamical systems are governed by a huge
number of unknown parameters and laws. Most of the time the
understanding of such systems seems beyond human reach. Yet, the
ability to predict from past observations their future trajectory with
an acceptable error is often of paramount importance.  Such past
observations form a time series and the main concern of this paper is
the following
\par\smallskip\noindent\hfil
\begin{tabular}{@{}ll@{}}
  \multicolumn{2}{@{}l@{}}{\sc time series forecasting problem}\\
  \underline{Input}:&real-valued time series $x_1,\ldots,x_N$
                      and a forecast horizon $h\in\bbbn_+$,\\
  \underline{Output}:& estimated future values
                       $\hat{x}_{N+1},\ldots,\hat{x}_{N+h}$.\\
\end{tabular}
\par\smallskip\noindent
Note that $N$ is not fixed here and is a part of the input. As an
example, one may think of daily recording of the number of births in a
country, starting from, say, 1980 until today. The forecasting
consists in predicting the number of births that will take place
tomorrow, or, more generally, each of next $h$ days, if the forecast
horizon expressed in the number of days ahead is $h$. In order to
assess the quality of a forecasting method, one has to repeat the
prediction every day over an extended period of time. One has then to
compare formerly predicted values with actually recorded
ones. 
Besides recurrent neural networks
\cite{hbb:rnn2021survey} and observable operator models
\cite{tj:oom2015}, singular spectrum analysis (SSA) \cite{gz:ssa2020}
provides one of the most promising forecasting frameworks. Indeed, the
experiments of \cite{kp:wlen2017} show that the SSA outperforms
standard statistical methods in the field such as ARIMA (see
e.g.\@\cite{ha:forecasting2023}).

This paper reports an experimental investigation of univariate time
series prediction using SSA related method, namely the vector
forecasting (see e.g.\@\cite{gz:ssa2020}). The SSA involves several
computational steps among which the vector forecasting may be seen as
a final one.  In its most basic version, the SSA has to be provided
with an additional information that may require an expert knowledge
about the observed phenomenon. Is it possible to reduce such an
additional input so that SSA forecasting could be applied to phenomena
of unknown dynamics by a user with no knowledge in the underlying
field nor in statistics\,?  In other words, is SSA forecasting
methodology ready to be implemented within decision support tools that
would be adopted by policy makers, farmers or in business
administration\,?

A partially positive answer to the latter question comes from the
availability of several software packages for the SSA, notably SSA-MTM
toolkit \cite{tcd:ssa-mtm} and the \textbf{R} library \textsf{Rssa}
\cite{koro:rssa,gkz:rssa2018}. Although for an end user, SSA-MTM
toolkit may be more suitable due to its graphical interface, the
choice of \textsf{Rssa} for experiments reported here has been
motivated by an intrinsic flexibility of a library. Nevertheless,
besides an input time series, both SSA-MTM toolkit and \textsf{Rssa}
must be provided with a \emph{window length} and a list of
\emph{components} to be grouped together for the
forecasting.\footnote{However, in \textsf{Rssa}, automated grouping is
  also available.}  However, finding the best such parameters is a
challenge, even for SSA specialists or experts of the observed
phenomenon. Moreover, with an inadequate choice, the forecasting
accuracy can drop dramatically, as confirmed by this study. Hopefully,
several authors discuss methods for inferring automatically from the
input times series those parameters. The present paper reports an
experimental study of forecast quality obtained using such methods.
Can those methods be included in decision-support tools which
automatically select suitable parameters for the SSA and compute the
required forecast\,? Concerning the window length, the present paper
brings a positive answer. The result of authors' investigation is
that, among very few existing methods, the one of
\cite{maETal:wlen2012} gives the best accuracy of forecasting, at
least for meteorological time series used in this
paper. Unfortunately, concerning the grouping, the accuracy of the
only known truly automated method (available in \textsf{Rssa} package
\cite{gkz:rssa2018}) is not always satisfactory. Although a suggestion
for improvement is given at the end of Sect.\nbs\ref{sec:res}, one of
the conclusions of the paper is that truly automated grouping
algorithms or viable heuristics need yet to be developed and
implemented.

In order to recall the importance of the length of the window and the
choice of components to group, the next section reviews the essential
SSA steps that lead to the vector forecasting further explained in
Sect.\nbs\ref{sec:forecast}. Sect.\nbs\ref{sec:window} and
\ref{sec:group} review a few methods for selecting SSA parameters.
The data sets used in the reported experiments are introduced in
Sect.\nbs\ref{sec:data} and the experiments are described in
Sect.\nbs\ref{sec:exp}. Their outcome is discussed in
Sect.\nbs\ref{sec:res}.
Throughout this paper, $[n]$ stands for $\{1,\ldots n\}$.

\section{SSA steps towards the vector forecasting}\label{sec:ssa}
The history of the SSA has been sketched in many papers and several books
(e.g.\@ in \cite{gz:ssa2020}) and it is omitted in the present article.

As its name suggests, the SSA is a spectral method using the singular
value decomposition \cite{ey1936} for the analysis of times series.
SSA based forecasting techniques are important extensions of the SSA
itself. The latter consists of the following steps:
\begin{enumerate}
\item the embedding of the input time series in a vector space,
\item the singular value decomposition (SVD) of the trajectory matrix,
\item obtaining elementary matrices and their Hankelizations,
\item the inverse embedding yielding the time series decomposition,
\item the grouping of the time series components.
\end{enumerate}
These steps are now explained.

\subsection{Embedding}\label{subs:embed}
The first step of the SSA is an \emph{embedding} of a real-valued input
time series $\tsx = (x_1,\ldots,x_N)$, $N>2$, into a vector space
spanned by a sequence of $K$ \emph{lagged vectors}
$X_1,\ldots,X_K\in\bbbr^L$, with
$X_i:=(x_i,x_{i+1},\ldots,x_{i+L-1})^\top$ where $L\in\bbbn$ is the
\emph{window length} and $K:=N-L+1$. Those lagged vectors form a
\emph{trajectory matrix} $\trx\in\bbbr^{L\times K}$,
$\trx=[X_1,\ldots,X_K]$,
\[
  \trx\ \ :=\quad\begin{pmatrix}
    x_1&x_2&x_3&\cdots&x_K\\
    x_2&x_3&x_4&\cdots&x_{K+1}\\
    x_3&x_4&x_5&\cdots&x_{K+2}\\
    \vdots&\vdots&\vdots&\ddots&\vdots\\
    x_L&x_{L+1}&x_{L+2}&\cdots&x_N
  \end{pmatrix}
\]
which is a Hankel matrix, viz., the elements of each anti-diagonal are
equal. The $k$-th anti-diagonal consists of those element of the matrix
that are indexed by $A_k$ defined in Eq.\nbs\eqref{eq:antidiag}
below. Note that this embedding is a bijection between the set of
sequences of length $N$ and the set of $L\times K$ Hankel
matrices. Consequently, by the \emph{inverse embedding}, from any
$m\times n$ Hankel matrix, one gets the corresponding sequence of
length $m+n-1$.

The parameter $L$ is essential for the whole SSA and is usually chosen
so that $L\le N/2$. This is assumed throughout the paper so as to keep
$L<K$. As the forecasting problem becomes trivial when the rank of
$\trx$ is strictly less than $L$, that special case is not considered
here in order to simplify the mathematical treatment. Indeed, when
the input time series comes from an intricate dynamical system,
as those used in this paper, one cannot expect that the rank
of $\trx$ is strictly less than $L$.

The above embedding $\tsx\mapsto\trx$ may be interpreted as a representation
of $\tsx$ by a trajectory of a hypothetical dynamical system that
generated $\tsx$.

\subsection{Singular value decomposition}
In the second step of the SSA, the singular value
decomposition (SVD) \cite{ey1936} of the trajectory matrix is computed:
$\trx = \bU\bSg\bV^\top$ where
$\bU = [U_1,\ldots, U_L]\in\bbbr^{L\times L}$ and
$\bV = [V_1,\ldots, V_K]\in\bbbr^{K\times K}$ are unitary matrices and,
$\bSg\in\bbbr^{L\times K}$ is a rectangular diagonal matrix with
diagonal $\sigma_1\ge\sigma_2\ge\cdots\ge\sigma_L\ge 0$, viz.,
$\Sigma_{kk}=\sigma_k$. Note that $\rnk(U)=L$ because $\rnk(\trx)=L$.

\subsection{Obtaining elementary matrices}
Every \emph{eigentriple}
$(U_k,\sigma_k,V_k)$, for $k\in[L]$, where $U_k$ (resp.\@ $V_k$) is
called \emph{left} (resp.\@ \emph{right}) \emph{singular vector} for
\emph{singular value} $\sigma_k$, yields an \emph{elementary matrix}
$\trx_k:=\sigma_kU_kV_k^\top$ of rank 1 so that
$\trx=\sum_{k=1}^L\trx_k$.  Note that all non null elementary matrices
in this decomposition are pairwise orthogonal.

By averaging over anti-diagonals, from an elementary matrix $\trx_k$
one gets its \emph{Hankelization} (see also \cite{sc:compl2021})
$\ttrx_k\in\bbbr^{L\times K}$\!. More precisely the $k$-th
anti-diagonal of an $L\times K$
matrix has its indexes in
\begin{equation}\label{eq:antidiag}
  A_k\quad:=\quad\{(i,j)\in[L]\times[K]:i+j=k+1\}\quad\text{for }k\in[N]
\end{equation}
and the Hankelization $\ttrx_k$ of $\trx_k$ is defined by
\[
  \tilde{x}_{k,i,j}\quad:=\quad\frac{1}{|A_{i+j-1}|}\sum_{p+q=i+j}x_{k,p,q}
\]
where $\tilde{x}_{k,i,j}$ (resp.\@ $x_{k,p,q}$) stands for the element
at row $i$ (resp.\@ $p$) and column $j$ (resp.\@ $q$) of $\ttrx_k$
(resp.\@ $\trx_k$).

\subsection{Inverse embedding}
As mentioned in Subsection\,\ref{subs:embed}, the embedding is a
bijection between the set of sequences of length $N$ and the set of
$L\times K$ Hankel matrices. Thus, on every Hankelized elementary
matrix $\ttrx_k$ one can apply the inverse embedding and obtain a
time series written $\tsx_k$ and called an \emph{elementary
  component} time series. The latter satisfy
$\tsx=\sum_{k=1}^L\tsx_k$.

\subsection{Grouping}
This step consists in defining a partition $\calJ\subseteq 2^{[L]}$
of $[L]$ so that each cluster $I\in\calJ$ determines a ``group''
$\tsx_I:=\sum_{k\in I}\tsx_k$ of elementary component time series
which is relevant for the input time series analysis. The relevance
of such groups is merely subjective. One may wish for instance
obtaining three groups reflecting respectively the trend, the
seasonality and the noise.

For the forecasting method used in this study, the grouping aims to
abstract from inessential details of the observed trajectory. Thus it
consists in choosing a strict subset $I$ of $[L]$ to get the
``signal'' $\tsx_I:=\sum_{k\in I}\tsx_k$ separated from the ``noise''
$\tsx_{\oI}:=\sum_{k\in[L]\setminus I}\tsx_k$. Similarly, one may
write $\trx$ as the sum of its relevant parts
$\trx_I:=\sum_{k\in I}\trx_k$ and its noisy part
$\trx_\oI:=\sum_{k\in[L]\setminus I}\trx_k$ or their Hankelizations
$\trx = \ttrx_I + \ttrx_\oI$. The reader should note that $I$ is
another additional input for the SSA and that the choice of $I$
greatly impacts the quality of forecasting \cite{kp:wlen2017}.

\section{Vector forecasting}\label{sec:forecast}
The vector forecasting is not a part of the SSA \textsl{per se}. Like two other
forecasting methods described in \cite{gz:ssa2020}, the vector forecasting
uses the SSA for extracting a low rank approximation of a subspace of a
hypothetical dynamical system that generated $\tsx$. The common idea
of the three forecasting methods presented in \cite{gz:ssa2020} is to
find a homogeneous linear recurrent equation (LRE)
\begin{equation}\label{eq:rec}
  x_{_{I\!,i+L-1}}\quad=\quad\sum_{k=1}^{L-1} a_kx_{_{I\!,i+L-1-k}}
  \quad\text{for}\ 1\le i\le K    
\end{equation}
that is satisfied by ``denoised'' time series
$\tsx_I=(x_{_{I\!,1}},\ldots,x_{_{I\!,N}})$. For $\tsx_I$ to
satisfy an LRE means it is generated by a linear dynamical system.
This lets $\tsx_I$ a wide class of behaviours. Although ``linearity''
may sound restrictive for a computer scientist, within the theory of
dynamical systems it refers to the underlying
evolution functions not to the behaviour of the system itself.
Indeed, it is well
known that linear dynamical systems behave like a sum of products of
polynomials, exponentials and sinusoids \cite{hesp:lin2018}.
Finding coefficients
\[
  \ba\quad:=\quad(a_{L-1},\ldots, a_1)
\]
of LRE~\eqref{eq:rec}
amounts to solving the following system of $K$ equations with
$L-1$ unknowns
\begin{equation}\label{eq:syst}
  \ba\utrx_I=(x_{_{I\!,L}},\ldots,x_{_{I\!,N}})
\end{equation}
where $\utrx_I$ denotes matrix $\trx_I$ without its last row
which in turn forms  the right hand side of the system of equations.
As $L\le N/2$ and hence $L<K$, this system is overdetermined, and in
general, has no exact solution, except when $\tsx_I$ is actually
governed by a linear dynamical system of dimension at most
$L$. However, as dynamical systems considered in this paper are not
linear, and this is also the case for all systems of intricate
dynamics, only approximate solutions of \eqref{eq:syst} can be
obtained.  This can be compared with a linear regression where one
fits a line to a set of points in an optimal way. In LRE-based
forecasting, such as the vector forecasting used in this paper, one
fits a linear recurrent sequence to the set of observations. For that,
the closest approximate solution of \eqref{eq:syst} with respect to
the Euclidean norm is sought.  It is well known that such an
approximate solution is given by
$\ba\approx (x_{_{I\!,L}},\ldots,x_{_{I\!,N}})\utrx_I^\dagger$, where
``$\_^\dagger$'' stands for the pseudo-inverse of Moore-Penrose of a
rectangular matrix \cite{penrose:approx1956}.

Let $\bU_I := [U_i: i\in I]$ be the matrix formed with columns of
$\bU$ with indexes in $I$ and let $\cU_I$ be the subspace of $\bbbr^L$
spanned by columns of $\bU_I$. Remember that $\bU$ is the matrix of
left singular vectors in the SVD of $\trx$.  Let $(u_{L,i}: i\in I)$
be the last row of $\bU_I$ and let $\ubU_I = [\uU_i: i\in I]$ stand
for $\bU_I$ with its last row removed. Similarly, let $\ucU_I$ be the
subspace of $\bbbr^{L-1}$ spanned by columns of $\ubU_I$.  Note that
$(u_{L,i}: i\in I)$ can be expressed as a linear combination of rows
of $\ubU_I$ because $\ubU_I$ is of rank $|I|<L$. Let
$\upsilon^2:=\sum_{i\in I}u_{L,i}^2$. Observe that
$\upsilon=\cos\theta$, where $\theta$ is the angle between $\cU_I$ and
$\be_L=(0,\dots,0,1)^\top\in\bbbr^L$. Indeed, the orthogonal projection
of $\be_L$ onto $\cU_I$ is expressed by
$\sum_{i\in I}(\be_L^\top U_i) U_I=\sum_{i\in I}u_{L,i}U_I$
and $\|\sum_{i\in I}u_{L,i}U_I\|^2=v^2$ because $\bU$ is unitary.
As $\be_L\notin\cU_I$, one has
$\upsilon\neq 1$ and the following vector is well defined
\[
  R\quad:=\quad\frac{1}{1-\upsilon^2}\sum_{i\in I}u_{L\!,i}\,\uU_i\enspace.
\]
It can be shown that
$(x_{_{I\!,L}},\ldots,x_{_{I\!,N}})\utrx_I^\dagger=R^\top$.
Consequently $R$ is the closest approximate solution of
\eqref{eq:syst}.  Now, matrix
\[
  \Pi\quad:=\quad\ubU_I\ubU_I^\top + (1-\upsilon^2)RR^\top
\]
defines the orthogonal projection of $\bbbr^{L-1}$ onto $\ucU_I$.  Let
$\check\bz$ be a vector $\bz\in\bbbr^L$ without its first coordinate.
Using a linear operator $\bF\colon\bbbr^L\to\cU_I$ which extends the
orthogonal projection $\Pi\check\bz$ of $\check\bz$ with the next term
of the recurrent sequence inferred from \eqref{eq:rec}
\[
  \bF\,\bz\quad:=\quad\left({\Pi\check\bz}\atop{R^\top\check\bz}\right)
\]
one defines a sequence of vectors
\[
  Y_i\quad:=\quad\left\{
    \begin{array}{@{}ll}
      X_{I,i}&\text{for }i\in[K],\\
      \bF Y_{i-1}&\text{for }i\in\{K+1,\ldots,N+h\},
    \end{array}\right.
\]
where $[X_{I,1},\ldots,X_{I,K}]=\trx_I$ and $h\in\bbbn$ is a forecast
horizon. This leads to matrix
\[
\bY\quad:=\quad[Y_1,\ldots,Y_{N+h}]
\]
obtained by extending $\trx_I$ on the right with
vectors $Y_{K+1},\ldots, Y_{N+h}$ resulting from iterating $\bF$ on $X_K$,
where $Y_{K+i}=\bF^iX_K$. In this context, operator $\bF$ can be understood
as a recurrence over vectors of $\cU_I$ obtained by an appropriate
lifting of LRE \eqref{eq:rec}.
By Hankelization of $\bY$ and its subsequent inverse embedding, one gets
a time series $\tsy=(y_1,\ldots,y_{N+h+L-1})$ where the portion
$(y_{N+1},\ldots,y_{N+h})$ is the forecast up to horizon $h$ obtained
by the vector forecasting method.

\section{Choice of the window length}\label{sec:window}
From the presentation of the SSA method including the vector forecasting,
it follows that the length of the window, $L$, is a crucial parameter.
Indeed, $L$ should be understood as the chosen dimension for the
model, built from the SSA, of the observed dynamical system.
It determines the order of LRE \eqref{eq:rec} which is precisely $L-1$.
Foundational texts (e.g. \cite{bk:dyna1986,vg:ssa1989}) and books
\cite{et:ssa1996,gnz:ssa2001,gz:ssa2020} give no general estimation
methods of this parameter. The prevailing opinion is that choosing $L$
only slightly less than $N/2$ allows capturing all significant
frequencies of periodic components of the underlying dynamical system.
Choosing $L$ equal to the longest oscillation period or a
multiple of that period not exceeding $N/2$ is also
often suggested. Unfortunately, if the data comes from a poorly
understood dynamical system, such a period is unknown.
Moreover, the outcome of experiments reported in
Sect.\,\ref{sec:res} suggests that such choices can result in
accuracies not much better than those of the random forecasting.
Therefore, providing methods and, more importantly, efficient
algorithms for estimating $L$ from the input time series is essential.

An appealing formal approach for estimating an adequate window length
is developed in \cite{kp:wlen2010} throughout an adaptation to the SSA of
the minimum description length principle (see e.g.\@
\cite{gr:mdl2010,gru:mdl2007}) better known as Kolmogorov
complexity. The method consist in a cross-optimisation of two
functions, say $f(L,M)$ and $g(L,M)$ wrt.\@ $L$ and $M$. This yields
an estimation of $L$ and also of the number $M$ of the most
significant components of $\tsx$ to be considered as signal.
Unfortunately, for each evaluation step of $f(L,M)$ or $g(L,M)$,
singular values of $\trx$ have to be computed because $\trx$ depends
on $L$. As a practical rule, the authors of \cite{kp:wlen2010} suggest
to take $(\log N)^c$ with $c\in(1.5,2.5)$ as an upper bound for $L$.
Although $(\log N)^c\in o(N)$ and therefore $(\log N)^c\ll N/2$, for
$N$ sufficiently large, with $c=2.5$ the method is still
computationally demanding when $\tsx$ is a time series with daily
samples over, say, 50 years, because the maximum value of $L$ then
equals $301$. Indeed, in case of an exhaustive search over
$L\in\{2,\ldots,301\}$, one has to perform $300$ singular value
decompositions (SVD).  Beyond formal demonstrations, in
\cite{kp:wlen2012} and \cite{kp:wlen2013} the authors of
\cite{kp:wlen2010} provide an experimental evaluation of their method
on real world data sets which confirms that choosing $L$ much smaller
than $N/2$ significantly improves the quality of forecasting.

Several authors use the \emph{autocorrelation function}
\begin{equation}\label{eq:autocorr}
  R(\tau):=\frac{1}{s^2}\sum_{i=1}^{N-\tau}(x_{i+\tau}-\ox)(x_i-\ox)
\end{equation}
where $\ox:=\frac{1}{N-\tau}\sum_{i=1}^{N-\tau}x_i$ (resp.\@
$s^2:=\frac{1}{N-\tau}\sum_{i=1}^{N-\tau}(x_i-\ox)^2$) is the
empirical mean (resp.\@ empirical variance) of $\tsx$.
In \cite{tpt:WLAN2009}, the smallest value of $\tau$ where $R$
crosses the confidence interval corresponding to ($95\%$ of) the white
Gaussian noise (with parameters $\ox$ and $s$) is used as estimate of
$L$.  In \cite{maETal:wlen2012} and \cite{wangETal:wlen2015} the
smallest value of $\tau$ such that $R(\tau)R(\tau+1)<0$ is used as
estimate of $L$.

In the sequel, $L^{\text{\cite{maETal:wlen2012}}}$ stands for the
length of the window chosen with the latter method whereas
$L_{\text{lo}}$ and $L_{\text{hi}}$ denote two extreme values for the
maximum window length in $\{(\log N)^c: c\in(1.5,2.5)\}$ discussed
formerly.

\section{Choice of the grouping}\label{sec:group}
The choice of index set $I$ of components that are used as signal in
forecasting is as essential as the choice of the window
length. Indeed, as mentioned earlier, the SSA should be understood
merely as a method for separating the true signal from the noise
within the raw signal obtained from observations.
Besides theoretical results, the experiments carried in
\cite{kp:wlen2013} show that both the grouping and the window length
selection have a tremendous impact on forecast accuracy. This is not a
surprise as both affect directly LRE \eqref{eq:rec}.

On the contrary to the window length where the search space is linear
in $N$ (yet brute force methods are limited by the computationally
costly step of SVD), the search space for grouping is in $O(2^L)$.
Several authors only consider groupings such that $I=[M]$ with
$M\in[L-1]$ which lets reducing the search space into $O(L)$.  In
other words, the signal is selected as the first $M$ elementary
components of $\tsx$, viz., $I:=[M]$ and $\tsx_I = \sum_{i=1}^M\tsx_i$.
This shall be called a \emph{prefix grouping} in the sequel.

A common practice for the grouping (see e.g.\@ \cite{gz:ssa2020}) is to
rely on visual examination of scatter plots and recurrence plots which
involves subjective assessment of parameters. Although, pattern
recognition techniques can be used within this approach, those also
require some parameters.

The \textbf{R} package \textsf{Rssa} implements two methods for
the grouping. The first method uses frequency analysis via discrete
Fourier transform. Again, as it requires additional parameters, it cannot be
qualified as automated grouping. The second method runs a clustering
algorithm using a similarity matrix between time series' elementary
components. That similarity matrix, $(s_{i,j})\in\bbbr^{L\times L}$,
which is more precisely a
\emph{$\bw$-correlation matrix}, is defined upon the following
\emph{weighted inner product}
\[
  (\tsy,\tsz)_\bw\quad:=\quad\sum_{i=1}^N|A_i|y_iz_i,
\]
where $\tsy$ and $\tsz$ are time series of length $N$,
and the corresponding \emph{weighted norm}
\[
  ||\tsx||_\bw:=\sqrt{(\tsx,\tsx)_\bw}\enspace.
\]
Remember that $A_i$,
defined in Sect.\nbs\ref{sec:ssa} Eq.\nbs\eqref{eq:antidiag} is the set of
indexes of the $i$-th anti-diagonal of an $L\times K$ matrix.
The $\bw$-correlation matrix is defined as follows
\[
  s_{i,j}\quad:=\quad\frac{(\tsx_i,\tsx_j)_\bw}{||\tsx_i||_\bw||\tsx_j||_\bw}
  \enspace.
\]
Function \textsf{grouping.auto.wcor} from \textsf{Rssa} implements the
latter clustering-based grouping. It can be considered as a fully
automated grouping method.

There is no \textsl{a priori} restriction on the form of index set $I$
of ``signal'' resulting from clustering by
\textsf{grouping.auto.wcor}.  On the contrary, the method of
\cite{kp:wlen2010} based on minimum description length yields
$M\in [L-1]$ to be used as prefix grouping $[M]$. Unfortunately, the
method is difficult to implement. No algorithm is clearly
stated. Implementations, if any, do not seem available in the public
domain.

\section{Data sets}\label{sec:data}
Several methods discussed above have been evaluated as a part of the
present work using real world data summarised in Table~\ref{tab:data}.
\begin{table}
  \begingroup\footnotesize
  \begin{tabular}{|l|l|l|l|l|l|}\hline
  \hfil\bfseries kind&\hfil\bfseries unit&\hfil\bfseries location&
  \hfil\bfseries coordinates&\hfil\bfseries begins on&\hfil\bfseries ends on\\\hline
  maximum temperature 24h&°C&Maevatanana&16°57'S 46°50'E
                            &1979-01-01&2017-12-31\\\hline
  minimum temperature 24h&°C&Ambatolampy&19°23'S 47°25'E
                            &1979-01-01&2018-12-31\\\hline
  rainfall 24h&mm&Marovoay&16°6'S 46°38'E
                            &1979-01-01&2017-12-31\\\hline
  water vapor&kg/m²&Ambovombe&25°10'S 46°05'E
                            &1979-01-01&2018-12-31\\\hline
  ozone&kg/m²&Antananarivo&18°56'S 47°31'E
                            &1979-01-01&2018-12-31\\\hline
  mean pressure&Pa&Grande Comore&11°55'S 43°25'E
                            &1979-01-01&2016-12-31\\\hline
\end{tabular}
\endgroup
\caption{Data summary}\label{tab:data}
\end{table}
The data sets used here have been downloaded from the ERA-Interim
archive of the European Centre for Medium-range Weather Forecasts
(ECMWF).  The ERA-Interim archives historical forecasts for horizons
from 0 to 240 hours. These forecasts consist of reanalysis data. The
meaning of ``reanalysis'' for horizon 0 is that the data either come
form observations or, if an observation is unavailable at a given
location, the corresponding value is interpolated using a
meteorological model. Whether a data set comes entirely from
observations or has some interpolated parts (due e.g. to a time outage
of a recording station) does not matter for this study, in view of a
high accuracy of interpolation of those specialised models.  On the
contrary, the location of each data set matters, as explained in the
sequel. It should be noted that all data sets used here are
``forecasts'' for horizon 0 which means that these are not predicted
but rather measured values or, exceptionally, interpolated ones.

It is worth noting that some informal recommendations about choices
of the window length and of the grouping have been evaluated in
several papers mainly on synthetic data. When carried out on
real world data, the evaluations invalidated the relevance of those
recommendations. The authors preference for meteorological data is
guided not only by the availability but also by the importance of
atmospheric and oceanic phenomena in many areas of life. The data
sets chosen for this paper concern such
phenomena at the northern extremity of the Mozambique Channel
(``Grande Comore'') and at several locations in Madagascar.  This
region of the Western Indian Ocean has a tropical climate experiencing
different micro-climates from part to part. In addition to the concern
to compare different climatic parameters, each location also has
particularities. Maevatanana, (resp.\@ Ambatolampy, Ambovombe,
Antananarivo), is the place reputed to be the hottest (resp.\@ the
coldest, the driest, the most polluted) on the island. Marovoay is an
area with high agricultural potential. The study of rainfall is
therefore as interesting as it is essential. The study of the
atmospheric pressure in Grande Comore is mainly motivated by
forecasting the trajectories of cyclones.

All recordings are daily and start from 1979-01-01. Both water vapor
and ozone are expressed in kg/m² representing their total amount in
a column extending from the surface of the Earth to the top of the
atmosphere.

\section{Experiments}\label{sec:exp}
The aim of numerical experiments reported in this paper was to asses
the quality of SSA forecasting from user's point of view for a short
duration with horizon $h\in[30]$. Here ``short duration'' is relative
to the length of the time series. In fact, meteorologist speak of medium-range
when $h\in[10]$ and long-range when the horizon exceeds 7 days
although these limits are not strict. Specialised meteorological
models have excellent accuracy of forecasts up to 5 days. The choice
for $h\in[30]$ is motivated by a potential future comparative study
of forecasting accuracies of specialised meteorological models vs.\@
general-purpose time series forecasting methods such as those from the SSA.

\begin{figure}[htb]\noindent
  \includegraphics[width=\linewidth]{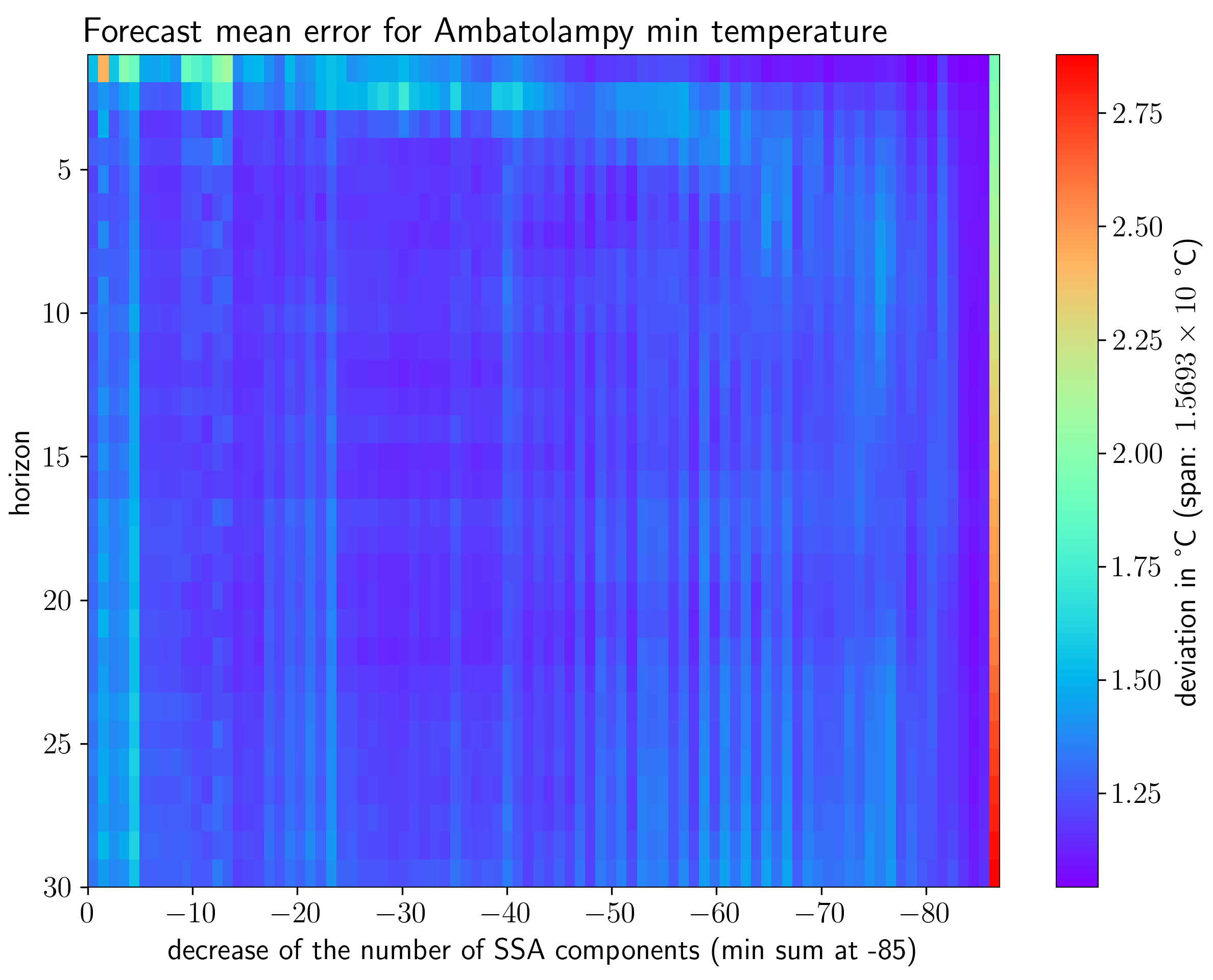}
  \caption{Forecast mean error for all prefix groupings from [89] down to [1]}
  \label{fig:ambato-color}
\end{figure}

For each data set, the forecast has been computed on every day of the
last year of data, except on December 31. For a given horizon $h$,
this resulted in $365-h$ forecasting days, except $366-h$ forecasting
days for leap year 2016 (Grande Comore time series only). Let $D_h$
(resp.\@ $F_h$) denote the set of forecasting (resp.\@ forecasted)
days for horizon $h$. For every forecasting day $j\in D_h$, computing
a forecast consisted in taking $\tsx_{\le j}:=(x_1,\ldots,x_j)$ as
input time series for
\begin{enumerate}
\item estimating the window length (see Sect.\nbs\ref{sec:window}),
\item embedding and decomposition (see Sect.\nbs\ref{sec:ssa}),
\item grouping (see Sect.\nbs\ref{sec:group}),
\item vector forecasting (see Sect.\nbs\ref{sec:forecast}),
\end{enumerate}
where the two latter steps were repeated using various grouping choices in
order to collect corresponding forecasts. Thus, for a fixed method of
the window length estimation, the most computationally expensive part,
namely SVD, was computed only once for each forecasting day.

\begin{figure}[htb]\noindent
  \includegraphics[width=\linewidth]{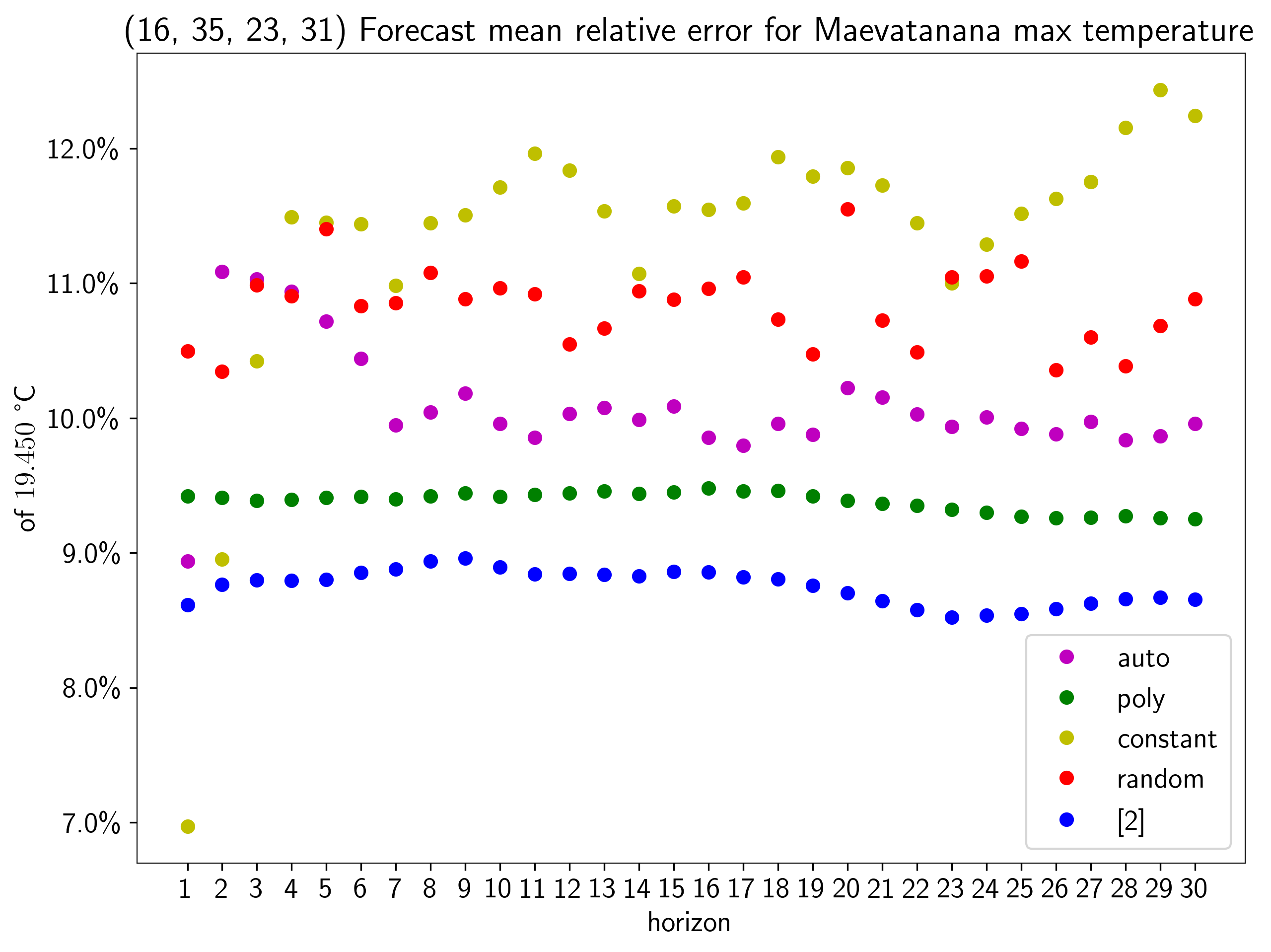}\hfil
  \caption{Forecast mean relative error for Maevatanana maximum temperature}
  \label{fig:maeva-mn}
\end{figure}

Assuming the methods for the window length and the grouping are fixed,
by repeating steps 1--4 for all forecasting days, one gets vectors
$(L_j:j\in D_1)$ and $(I_j:j\in D_1)$ of window lengths and groupings,
and, for every horizon $h$, a vector of forecasted values
\[
\by_h=(y_{h,j}:j\in F_h),
\]
where $y_{h,j}$ is the value for day $j+h$
forecasted from $\tsx_{\le j}$ (as if it were done on day $j$).
The \emph{forecasting error vector for
  $h$}, is therefore $\bxi_h:=\by_h-\bx_h$, where
$\bx_h=(x_j:j\in F_h)$ is the vector of the corresponding actual
values, viz., the corresponding terminal portion of $\tsx$. The
\emph{mean} (resp.\@ \emph{maximum}) \emph{error for} $h$ is
$\mean(|\bxi_h|)$ (resp.\@ $\max(|\bxi_h|)$) where ``mean'' stands for
the arithmetic mean. These absolute errors have their relative
variants, each one defined as the ratio of the corresponding absolute
error divided by span of the data, namely
\[
  \frac{\mean(|\bxi_h|)}{\max(\tsx)-\min(\tsx)}\quad\text{and}\quad
  \frac{\max(|\bxi_h|)}{\max(\tsx)-\min(\tsx)}\enspace.
\]
Although when it comes to forecasting, the mean squared error is mostly
used, the authors believe that the arithmetic mean has a clear
intuitive meaning for an average user. By the way, the maximum error
can also be important in many forecasting tasks. It can for instance
bring some insight about the ability to forecast extreme events. This
is particularly important for phenomena expressed by data sets used in
this study.

For comparison with SSA vector forecasting, the following ``naive''
forecasting methods have been used as benchmark at every forecasting day
$j\in D_h$:
\begin{itemize}
\item \underline{random forecast} -- the forecasted value is sampled
  from the distribution inferred from $\tsx_{\le j}$,
\item \underline{constant forecast} -- the forecasted value equals $x_j$,
\item \underline{regression based forecast} -- uses polynomial
  regression (with polynomials of degree 4)
  from $\tsx_{\le j}$ to extrapolate the value used as the forecast.
\end{itemize}

Only rough evaluation of forecast quality
\begin{table}[htb]\noindent\hfil
  \begin{tabular}{|*{7}{c|}}\cline{2-7}
    \multicolumn{1}{c|}{}&&&&&&\\
    \multicolumn{1}{c|}{}&$\displaystyle\min_{j\in D_1}L_{\text{lo},j}$&
         $\displaystyle\max_{j\in D_1}L_{\text{lo},j}$&
         $\displaystyle\min_{j\in D_1}L^{\text{\cite{maETal:wlen2012}}}_j$&
         $\displaystyle\max_{j\in D_1}L^{\text{\cite{maETal:wlen2012}}}_j$&                      $\displaystyle\min_{j\in D_1}L_{\text{hi},j}$&                                      $\displaystyle\max_{j\in D_1}L_{\text{hi},j}$\\[3ex]\hline
    Maevatanana&29&30&52&52&281&283\\\hline
    Ambatolampy&30&30&90&91&283&285\\\hline
    Marovoay&29&30&78&80&281&283\\\hline
    Ambovombe&30&30&90&91&283&285\\\hline
    Antananarivo&30&30&95&97&283&285\\\hline
    Grande Comore&29&29&90&90&279&281\\\hline
    \multicolumn{7}{c}{}\\
  \end{tabular}
  \caption{Window lengths computed using three methods}\label{tab:win}
\end{table}
with varying window lengths (see Table~\ref{tab:win})
has been conducted because of an important computational cost of the
decomposition step. On an \textsl{ad hoc} basis, a part of this
evaluation was done for window length $L_{\text{big},j}$ taken as the largest
multiple of a mean year duration $365.25$ smaller than $j/2$,
together with prefix grouping $I_{\text{big},j}=[M_{\text{big},j}]$
for $M_{\text{big},j}=L_{\text{big},j}-1$. Another part was done for
$L_{\text{hi},j}$ together with prefix grouping
$I_{\text{hi},j}=[M_{\text{hi},j}]$ for
$M_{\text{hi},j}=L_{\text{hi},j}-1$. Remember that
$L_{\text{lo},j}=\lceil(\log j)^{1.5}\rceil$ and
$L_{\text{hi},j}=\lfloor(\log j)^{2.5}\rfloor$ are extreme integer values
of possible window sizes suggested in \cite{kp:wlen2010}.

A deeper evaluation relied essentially on the method of
\cite{maETal:wlen2012} with estimating the window length for
$\tsx_{\le j}$ written $L^{\text{\cite{maETal:wlen2012}}}_j$.  In a
more systematic way, for each data set, and every forecasting day,
$L^{\text{\cite{maETal:wlen2012}}}_j$, $L_{\text{lo},j}$ and
$L_{\text{hi},j}$ have been computed but only
$L^{\text{\cite{maETal:wlen2012}}}_j$ and $L_{\text{lo},j}$ have been
used for forecasting with various groupings. More precisely, prefix
groupings have been performed for all
$M_j\in[L^{\text{\cite{maETal:wlen2012}}}_j]$ (resp.\@
$M_j\in[L_{\text{lo},j}]$).  Fig.\nbs\ref{fig:ambato-color} displays a
result of such an exhaustive evaluation.  By averaging over the last
year of the time series, the best \textsl{a posteriori} prefix
grouping $I_{\mean}=[M_{\mean}]$ (resp.\@ $I_{\max}=[M_{\max}]$) with
respect to the mean (resp.\@ maximum) forecast error has been selected
for comparing with automated groupings $I_{\text{auto},j}$ computed by
\textsf{grouping.auto.wcor}.  Also the closest neighbourhood
$\calV_{\mean}$ (resp.\@ $\calV_{\max}$) of $I_{\mean}$ (resp.\@
$I_{\max}$) has been examined. This neighbourhood
consists of all index sets that differ from $I_{\mean}$
(resp.\@ $I_{\max}$) by one element only:
\begin{align}\label{eq:neigh}
  \calV_{\mean}:=&\bigl\{[M_{\mean}]\setminus\{k\}:k\in[M_{\mean}]\bigr\}\cup{}\\
      &\bigl\{[M_{\mean}]\cup\{k\}:k\in[\displaystyle
        \min_{j\in D_1}L^{\text{\cite{maETal:wlen2012}}}_j]
                      \setminus[M_{\mean}]\bigr\}\nonumber\\
    \bigl(\text{resp. }
    \calV_{\max}:=&\bigl\{[M_{\max}]\setminus\{k\}:
                       k\in[M_{\max}]\bigr\}\cup{}\nonumber\\
    &\bigl\{[M_{\max}]\cup\{k\}:k\in[\displaystyle
        \min_{j\in D_1}L^{\text{\cite{maETal:wlen2012}}}_j]
                      \setminus[M_{\max}]\bigr\}\bigr)\nonumber
\end{align}

All numerical evaluations were programmed in Python and R.
The programs are available upon request from the corresponding author.

\begin{figure}[htb]\noindent
  \includegraphics[width=\linewidth]{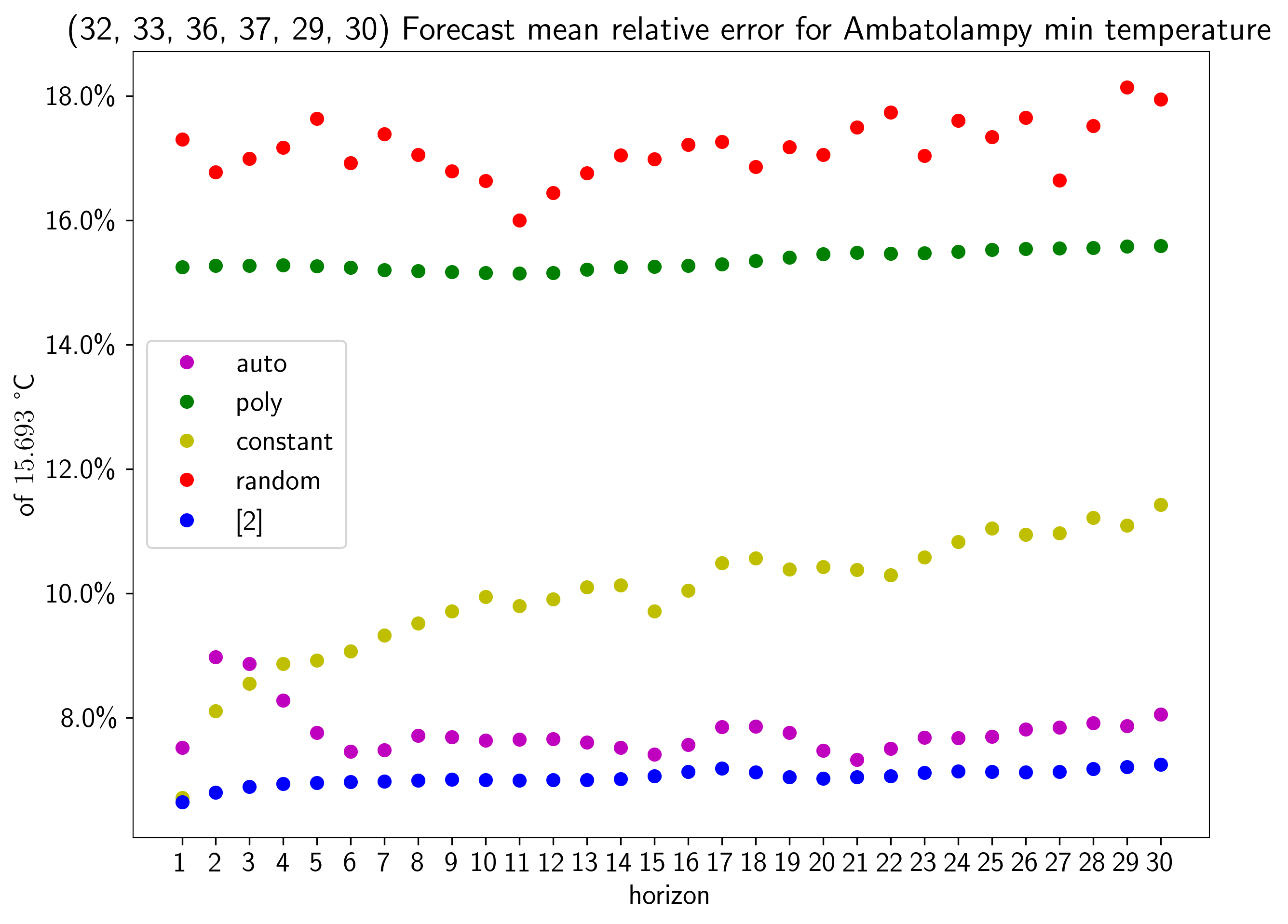}
  \caption{Forecast mean relative error for Ambatolampy minimum temperature}
  \label{fig:ambato-mn}
\end{figure}

\section{Results and discussion}\label{sec:res}
As a preamble to this section, it is worth mentioning
that the SSA and the corresponding forecasting methods are not
considered as learning algorithms. Nevertheless, some analogies
with machine learning or their lack are worth highlighting.
\begin{enumerate}
\item \underline{Undertraining}\ happens exactly as in machine
  learning when the input time series is too short to capture all
  essential behaviour of the observed dynamical system. In other
  words, there is not enough of observations.
\item \underline{Overfitting}\ results from the choice of index
  set $I$ including too many inessential components (obtained
  using SVD). This choice, called ``grouping'' is discussed in
  Sect.\nbs\ref{sec:group}. Since those inessential components are
  considered as noise, a model (an LRE in the case of SSA forecasting)
  capturing such noise is overfitted.
\item \underline{Underfitting}\ is, as usual, the opposite of overfitting.
  It occurs when index set $I$ does not include enough of essential
  components. The reader should keep in mind that $I\subseteq[L]$.
  Thus, when $L$ is too small, the underfitting
  cannot be compensated by a good choice of $I$.
\item The choice of window length $L$ does not seem to have a
  straightforward machine learning counterpart. It can be understood
  as the choice of the dimension of the model. This is for instance
  similar to the choice of the number of states of a hidden Markov
  model (see e.g.\nbs\cite{rabiner:HMM}) required by spectral learning
  algorithms \cite{hkz:spectralHMM2012} or by the classical Baum-Welch
  algorithm \cite{baum:baum-welch,welch:baum-welch}. The choice of $L$
  can be also compared to the choice of the degree of the polynomial
  in polynomial regression. Choosing this degree too big leads
  typically to an overfitting. The potential overfitting when
  $L$ is too big can be however compensated to some extent by
  the choice of $I$.
\end{enumerate}

\begin{figure}[htb]\noindent
  \includegraphics[width=\linewidth]{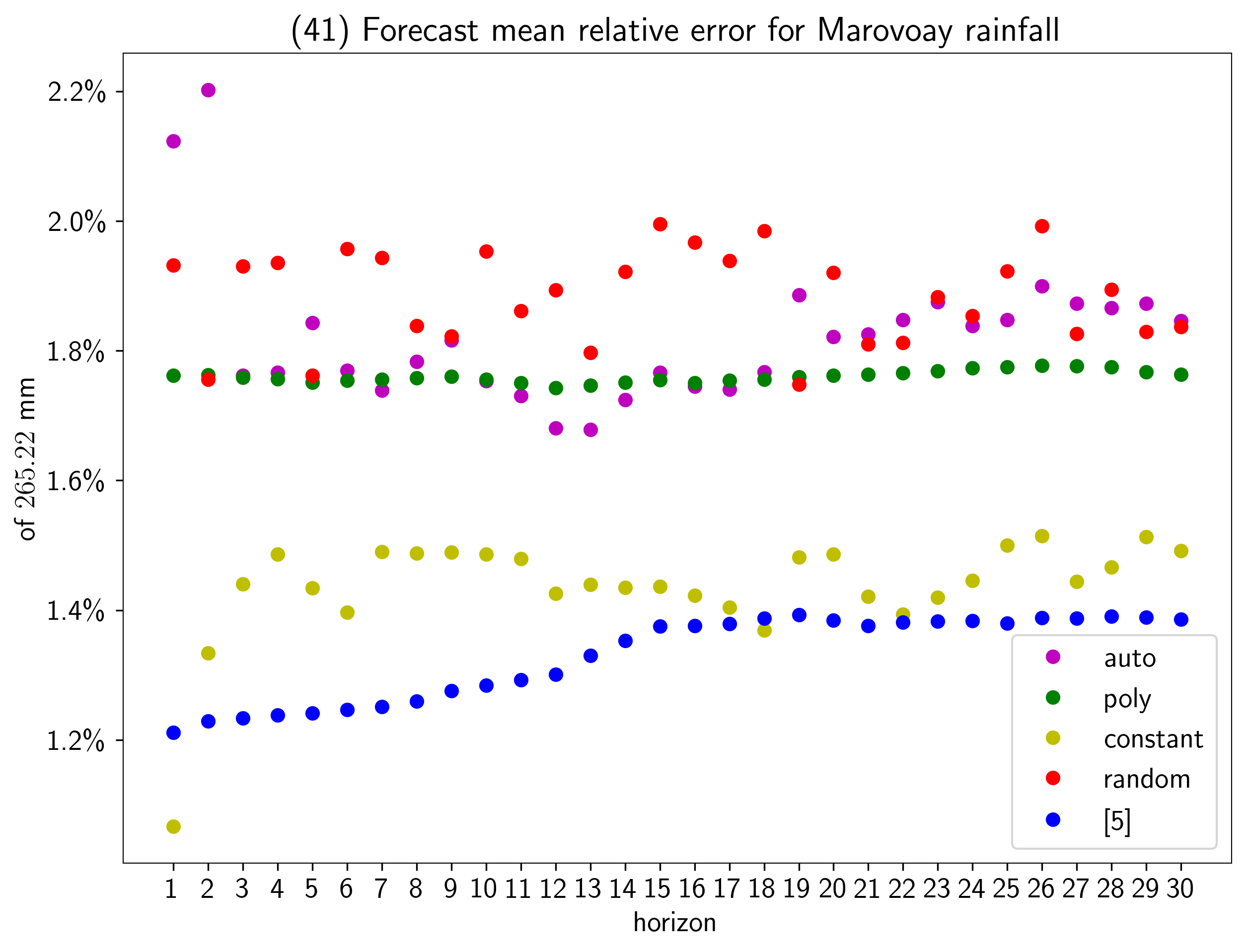}
  \caption{Forecast mean relative error for Marovoay rainfall}
  \label{fig:marovo-mn}
\end{figure}

While varying day $j$ over $D_1$, computed window lengths
$L_{\text{lo},j}$, $ L_{\text{hi},j}$. and $L^{\text{\cite{maETal:wlen2012}}}_j$
remained stable (see Table\nbs\ref{tab:win}).
Interestingly, for every considered data set and each $j\in D_1$,
one has
\[
  L_{\text{lo},j}<L^{\text{\cite{maETal:wlen2012}}}_j< L_{\text{hi},j}\enspace.
\]
This is surprising as $L_{\text{lo},j}$ and $L_{\text{hi},j}$
are the extreme integer values within the real interval
$[(\log j)^{1.5},(\log j)^{2.5}]$ which depends only on length
$j$ of $\tsx_{\le j}$. On the other hand, $L^{\text{\cite{maETal:wlen2012}}}_j$
depends not only on $j$ but also on the autocorrelation
function (see Eq.\nbs\ref{eq:autocorr}). Consequently, on the
contrary to $L_{\text{lo},j}$ and $L_{\text{hi},j}$,
$L^{\text{\cite{maETal:wlen2012}}}_j$ depends on the values of $\tsx_{\le j}$.
The above inequality is not a general rule and could be attributed to a
specific nature of considered data sets, all resulting from
atmospheric and oceanic phenomena.

A rough evaluation of forecasting with parameters
$(L_{\text{big},j}, I_{\text{big},j})$ and
$(L_{\text{hi},j}, I_{\text{hi},j})$ confirms the findings of
\cite{kp:wlen2010,kp:wlen2012,kp:wlen2013} and \cite{kp:wlen2017} that
using longer windows do not improve forecast accuracy. Indeed, the
forecasts obtained within the present experimentation with parameters
$(L_{\text{big},j}, I_{\text{big},j})$ were not better than random
ones. Even with $(L_{\text{hi},j}, I_{\text{hi},j})$ the accuracy was
only slightly better than using random forecast. The only convincing
results come with $L^{\text{\cite{maETal:wlen2012}}}_j$ and
$L_{\text{lo},j}$. These are depicted in Appendix
\ref{app:compar}. Their comparison lets conclude that the results
obtained with $L^{\text{\cite{maETal:wlen2012}}}_j$ are better than
with $L_{\text{lo},j}$.  Consequently the method of
\cite{maETal:wlen2012} based on the autocorrelation function (see
Eq.\nbs\eqref{eq:autocorr}) is favoured by the authors as it is easy
to implement and gives satisfactory results.

\begin{figure}[htb]\noindent
  \includegraphics[width=\linewidth]{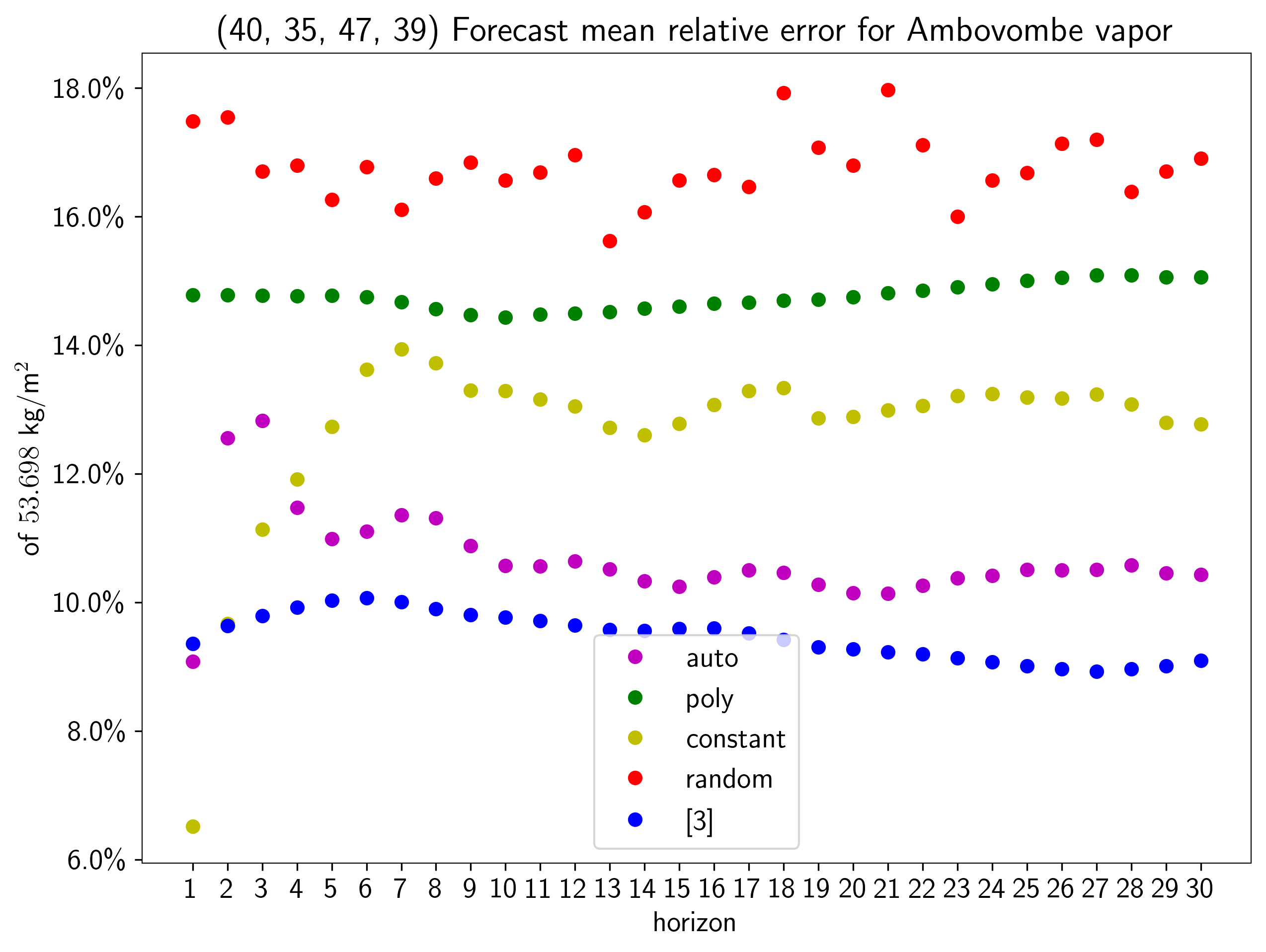}
  \caption{Forecast mean relative error for Ambovombe vapor}\label{fig:ambov-mn}
\end{figure}

A systematic evaluation of prefix grouping for window length
$L^{\text{\cite{maETal:wlen2012}}}_j$ shows that the
accuracy of forecasting is very sensitive to it.  One could think that
including more components would better capture the dynamics via an LRE
but often the opposite is true. Fig.\nbs\ref{fig:ambato-color} shows
how, in prefix grouping, decreasing the number of components affects
the forecast. The groupings considered in this example range from
$[88]$ down to $[1]$. One can observe that the mean error does not
show any regular pattern.  The best accuracy is obtained with $I=[2]$
when comparing mean errors averaged over all horizons from 1 to 30.

In all time series studied here, the optimal prefix grouping is
compared with with clustering-based grouping computed by function
\textsf{grouping.auto.wcor}. This is displayed on
Fig.\nbs\ref{fig:maeva-mn} to \ref{fig:comore-mn} which show
the mean error per horizon. Similar plots for maximum error
appear in Appendix \ref{app:max} and use the same colour codes.
The
errors are plotted in blue for optimum prefix grouping, in purple for
the automated grouping, in green for polynomial regression based
forecast, in yellow for constant forecast and in red for random
forecast. All errors are given relatively to the span
$\max(\tsx)-\min(\tsx)$ of the data set as precised on the left edge
of each plot.
A surprising observation common to all data sets discussed here is that
\textsf{grouping.auto.wcor} always returned a prefix grouping. This is
not a general rule. Indeed, for other time series,
\textsf{grouping.auto.wcor} may return a cluster of non prefix
groupings. Every plot on Fig.\nbs\ref{fig:maeva-mn} to \ref{fig:comore-mn} has
on top the list of groupings, in parentheses, computed by
\textsf{grouping.auto.wcor} when varying $j\in D_1$. More precisely,
an integer $k$ appearing on the list, means that for some $j\in D_1$,
\textsf{grouping.auto.wcor} returned index set $[k]$ after taking
$\tsx_{\le j}$ as the input time series. It should be noted here
that across $D_1$, very few different grouping are obtained and that
their variation is non-monotonic. As for all plotted
forecast errors for horizon $h$, the one resulting from automated
grouping is obtained by averaging over $j\in D_h$. When the mean error
is concerned (Fig.\nbs\ref{fig:maeva-mn} to \ref{fig:comore-mn})
the forecasts using automated groupings computed
by \textsf{grouping.auto.wcor} clearly appear as sub-optimal.  For
Marovoay rainfall (Fig.\nbs\ref{fig:marovo-mn}), that forecast is even
close to random forecast and for Maevatanana maximum temperature
(Fig.\nbs\ref{fig:maeva-mn}) a polynomial regression predicts more
accurately.

\begin{figure}[htb]\noindent
  \includegraphics[width=\linewidth]{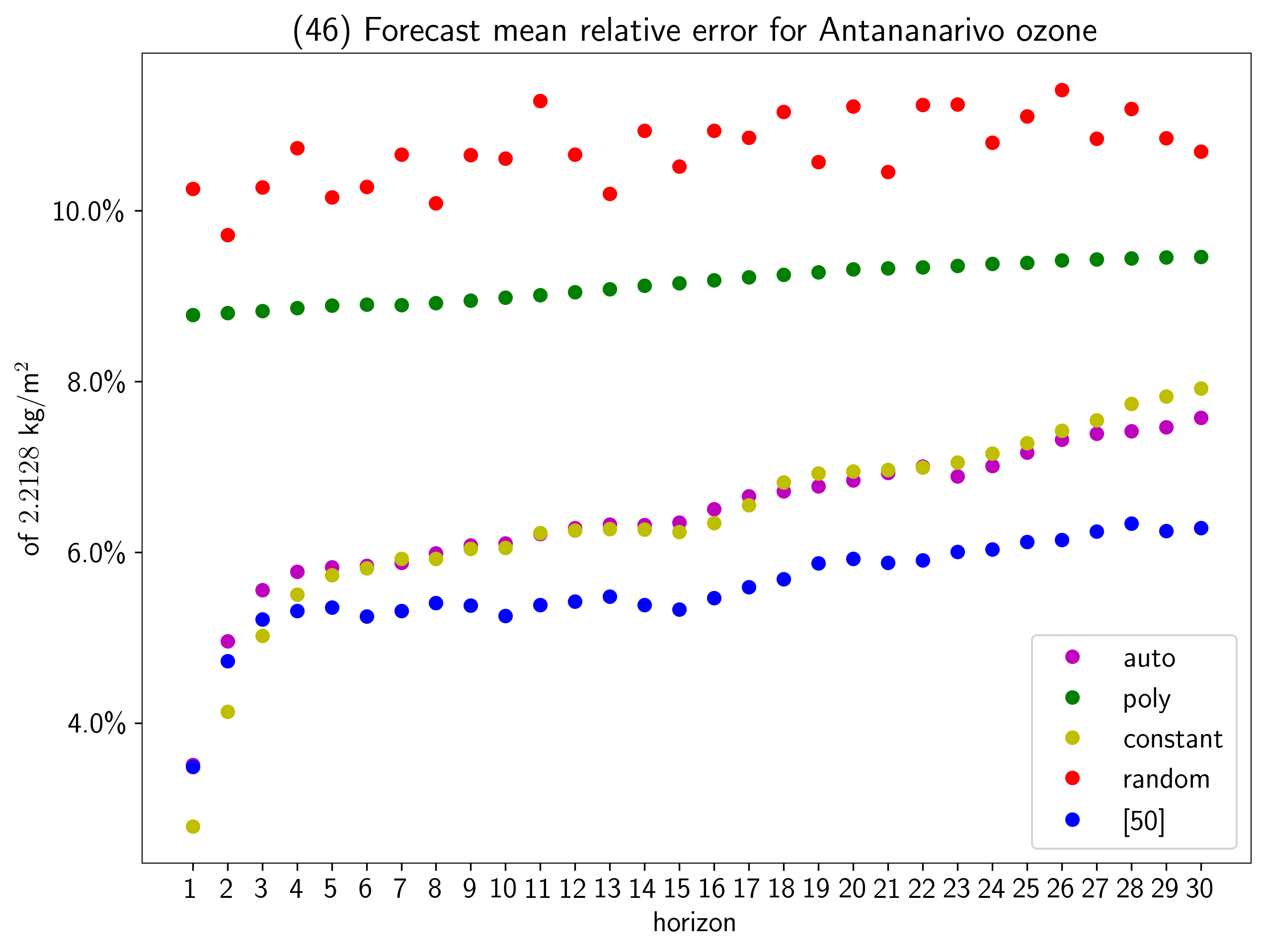}
  \caption{Forecast mean relative error for Antananarivo ozone}
  \label{fig:analam-mn}
\end{figure}

A systematic examination of prefix groupings for $L_{\text{lo}}$
confirms the above observations about automated grouping. It also
lets comparing forecast accuracy for window lengths
$L^{\text{\cite{maETal:wlen2012}}}$ and $L_{\text{lo}}$ (see figures in
Appendix \ref{app:compar}). As far as
mean errors are used for comparison, the automated (resp.\@
optimal) grouping
with $L_{\text{lo}}$ underperforms the automated (resp.\@
optimal) grouping with $L^{\text{\cite{maETal:wlen2012}}}$ in all
examples studied, except for Marovoay rainfall. However, when
one uses maximum errors (right column) instead of mean errors
(left column), no winner
can be clearly declared. Moreover, the plots of maximum errors for
$L^{\text{\cite{maETal:wlen2012}}}$
(see Appendix \ref{app:max})
seem to show that the SSA is unsuitable for forecasting
when maximum errors are the main concern. Indeed, even with
optimal prefix grouping, the maximum error is mostly beyond
$30\%$. This is perhaps a general drawback of all general-purpose
forecasting methods for time series, as the maximum error criterion
seems to be deliberately avoided in the corresponding literature.

\begin{figure}[htb]\noindent
  \includegraphics[width=\linewidth]{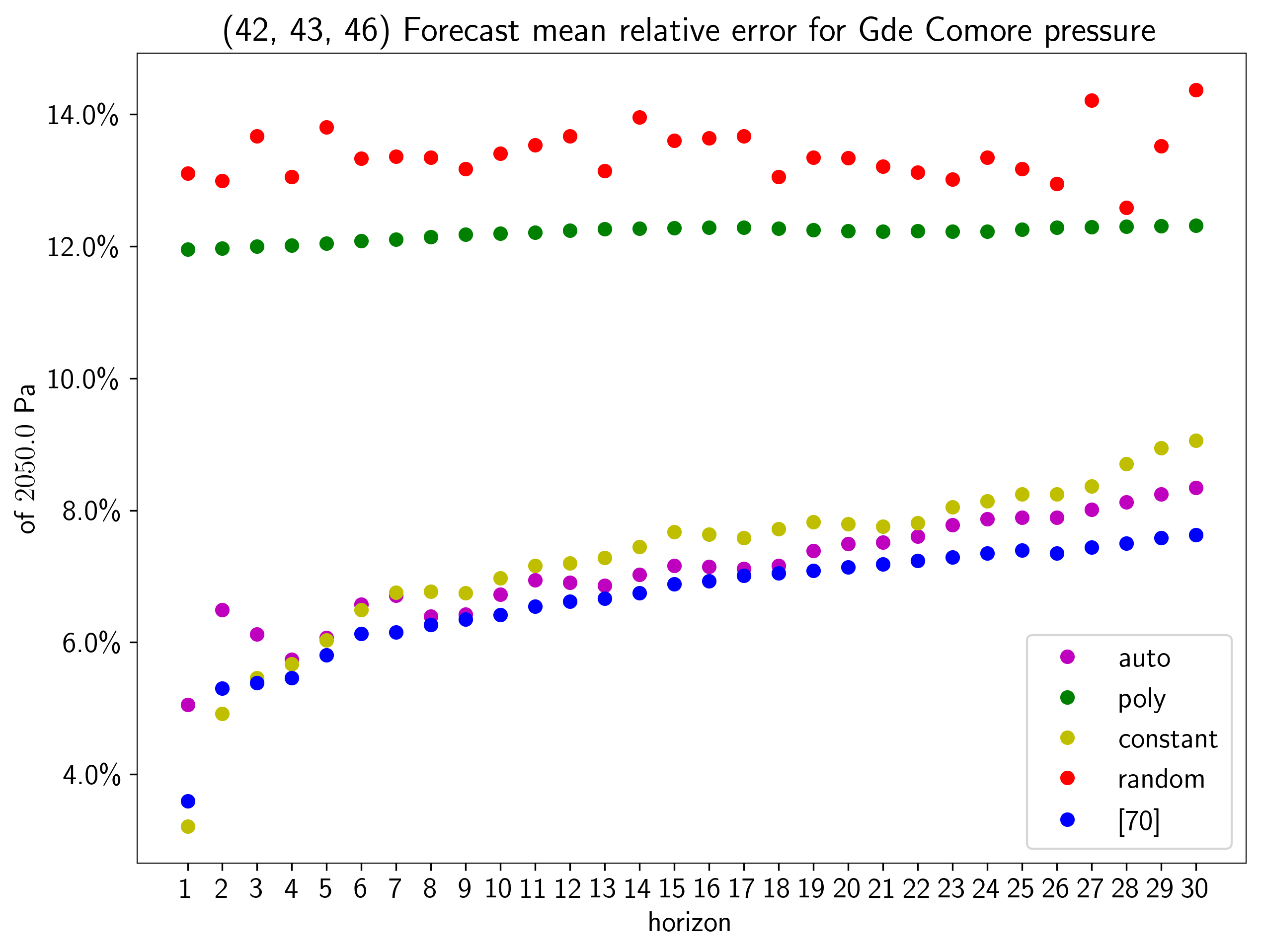}
  \caption{Forecast mean relative error for Grande Comore atmospheric pressure}
  \label{fig:comore-mn}
\end{figure}

As all groupings discussed in this section are prefix ones, one may ask
if, for a given window size, the optimal prefix grouping remains
optimal among all groupings. The authors do not know the answer
although they observed that the values of the errors in the closest
neighbourhoods $\calV_{\mean}$ or $\calV_{\max}$ (see
Eq.\nbs\eqref{eq:neigh}) of each optimal prefix groupings
(plotted in blue on all figures except on
Fig.\nbs\ref{fig:ambato-color}) always exceed those of the
latter. Therefore, each optimum prefix grouping for data sets
considered here form a local minimum. The authors do not know whether
this observation could be turned into a theorem nor if such local
minima are also global ones. In any case, the latter observation
leads to the conclusion that a viable strategy for
improving the automated grouping would be to start with the value
yield by \textsf{grouping.auto.wcor} and find a nearest local optimum
for prefix grouping.
The authors believe that similar results to theirs would be obtained
from other real world data, as long as they come from systems of
intricate dynamics and from sufficiently long time series.

\section{Conclusion}
The experiments reported in this paper confirm that the choice of the
window length and of the grouping are essential for the accuracy of SSA
forecasting. The window length selection method of
\cite{maETal:wlen2012} together with an adequate grouping enables
forecasting with an accuracy significantly better than constant or
random forecasting, provided that the mean error is
considered. However, the reader should keep in mind that each adequate
(optimal) grouping has been selected via an \textsl{a posteriori}
evaluation.  The only widely available method for an automated
\textsl{a priori} grouping, namely function
\textsf{grouping.auto.wcor} from \textsf{Rssa} package, appears as
sub-optimal in the analysed examples. Consequently, this is where the
research on the SSA should focus in order to make SSA forecasting ready
to be included in decision-support tools.
In the meantime, the results of this study suggest the following
strategy for a practitioner.
\begin{enumerate}
\item For every forecasting day, use a fixed-size suffix of the available
  time series as test data.
\item Perform a local search of a minarg, say $M_G$, around the
  set, say $G$, returned by \textsf{grouping.auto.wcor} upon
  applying SSA forecasting on the training data (viz., the portion of
  the time series without the test suffix) using test data for
  evaluating the error, say $e_{M_G}$.
\item Similarly, perform a local search for minarg, say $M_{[2]}$,
  around prefix grouping $\{1,2\}$. Let $e_{M_{[2]}}$ be the
  corresponding evaluation of the error.
\item For today's forecast, perform SSA steps up to grouping using
  the whole available time series.
\item If $e_{M_{[2]}}\le e_{M_G}$, use group $M_{[2]}$ for computing
  today's forecast.
\item If $e_{M_{[2]}}>e_{M_G}$, let $G'$ be the group returned by
  \textsf{grouping.auto.wcor}. For today's forecast use group
  $M_{G'}$ obtained from $G'$ by translation analogous to that
  from $G$ to $M_G$.
\end{enumerate}

The results also suggest a preference of the window length selection
method of \cite{maETal:wlen2012} -- it outperforms other methods
evaluated in reported experiments.  If however an implementation of
the method of \cite{kp:wlen2010} is available, it could be
advantageously used, as it simultaneously selects both the window
length and the grouping.

When the maximum error matters, SSA forecasting seems rather
unsuitable for short horizon prediction, at least for
atmospheric/oceanic phenomena illustrated by the time series used in
this study. The lack of literature with the maximum error criterion
does not let suggest an alternative.

When examining plots of forecasting errors using various methods,
one could ask, what could be learned from those about the dynamics of the
underlying phenomena. Somewhat intriguing is the fact that the
relative position of the plots differs substantially form one
time series to another. How to explain
that among ``naive'' forecasting methods the constant one is the
worst for Ambatolampy minimum temperature (see Fig\nbs\ref{fig:ambato-mn})
but the best one for Maevatanana maximum temperature
(see Fig\nbs\ref{fig:maeva-mn})\,?

\section*{Acknowledgements}
The authors wish to thank Juan Bógalo for general explanation about
the circulant SSA and Hong-Guang Ma for clarification about his method for
estimating the window length, and both, for providing their
Matlab codes.

\printbibheading
\printbibliography[heading=subbibliography,nottype=software,nottype=softwareversion,nottype=softwaremodule,nottype=codefragment,title={Publications}]
\printbibliography[heading=subbibliography,type=software,title={Software Project}]
\appendix
\section*{A p p e n d i c e s}
\section{Maximum error plots for \boldmath{$L^{[15]}$}}
\label{app:max}
\noindent
\includegraphics[width=.5\linewidth]{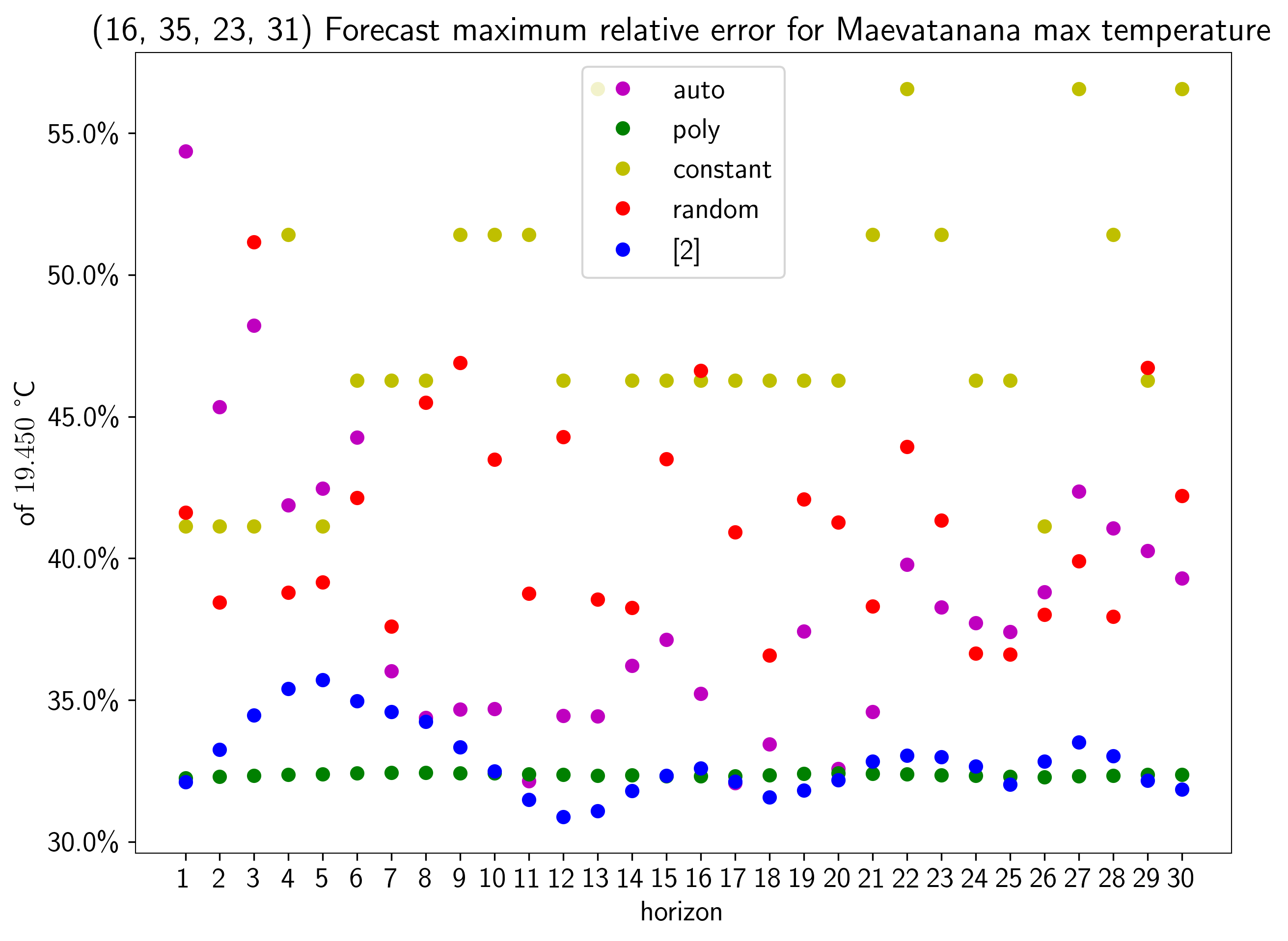}
\includegraphics[width=.5\linewidth]{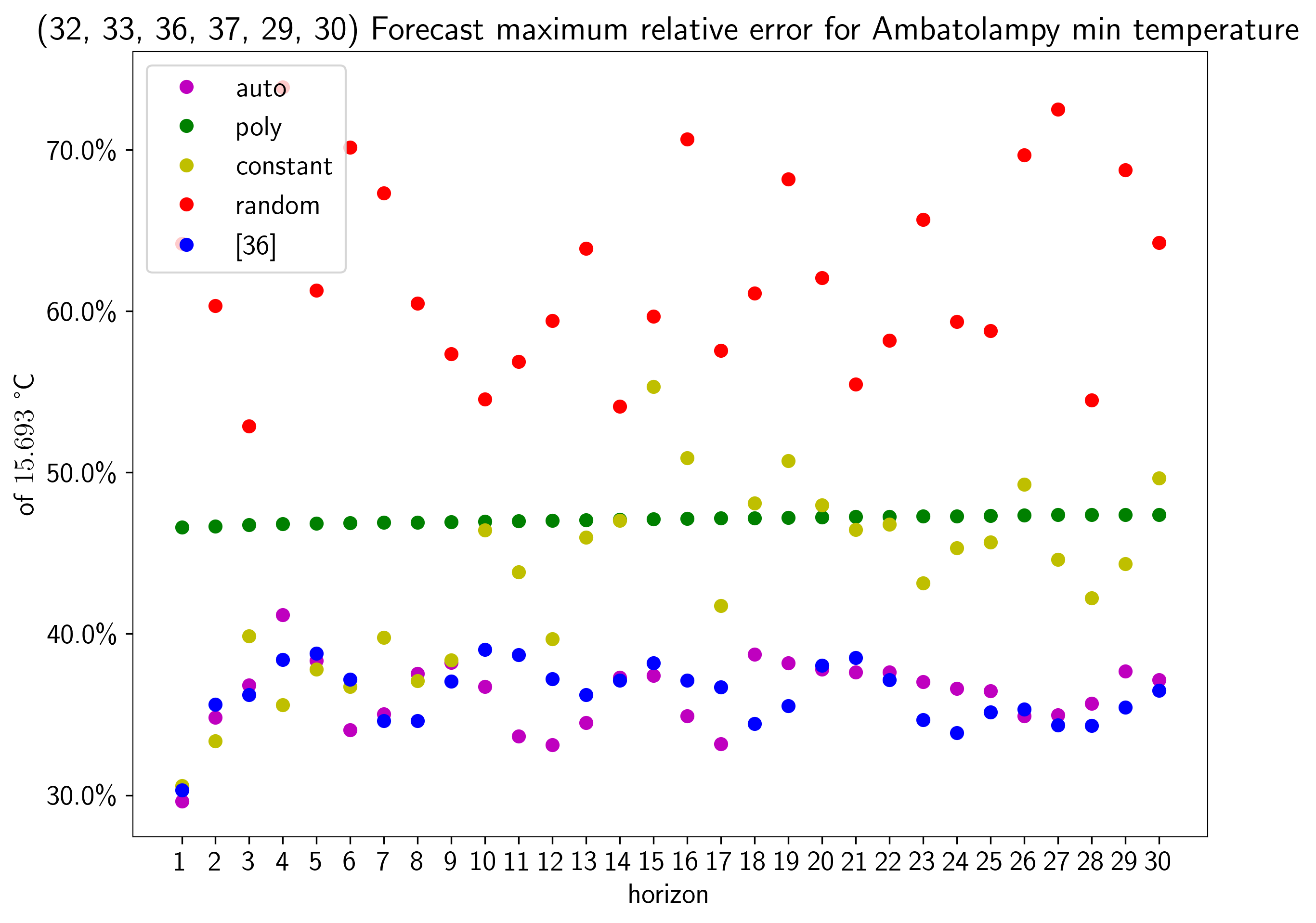}
\includegraphics[width=.5\linewidth]{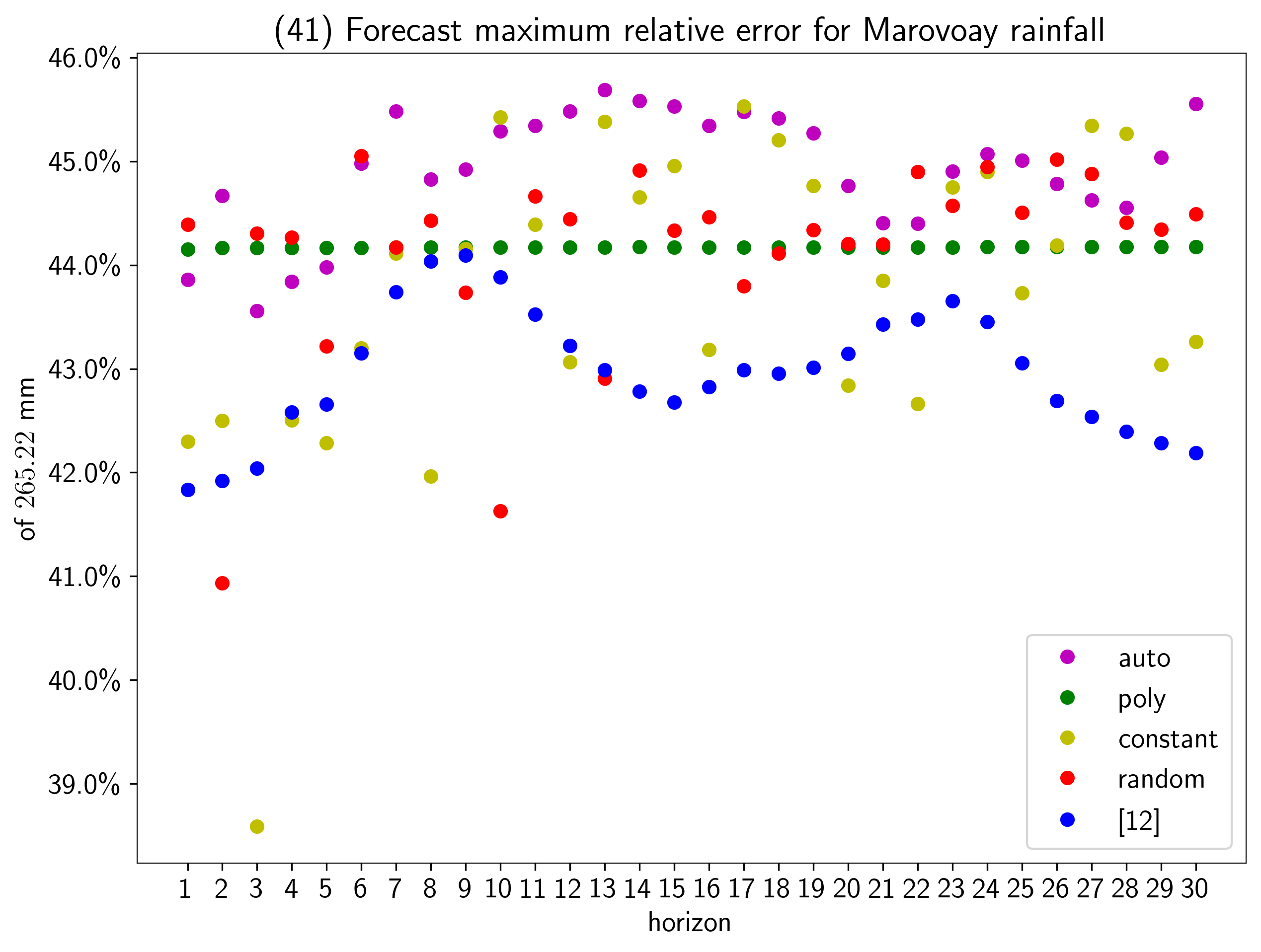}
\includegraphics[width=.5\linewidth]{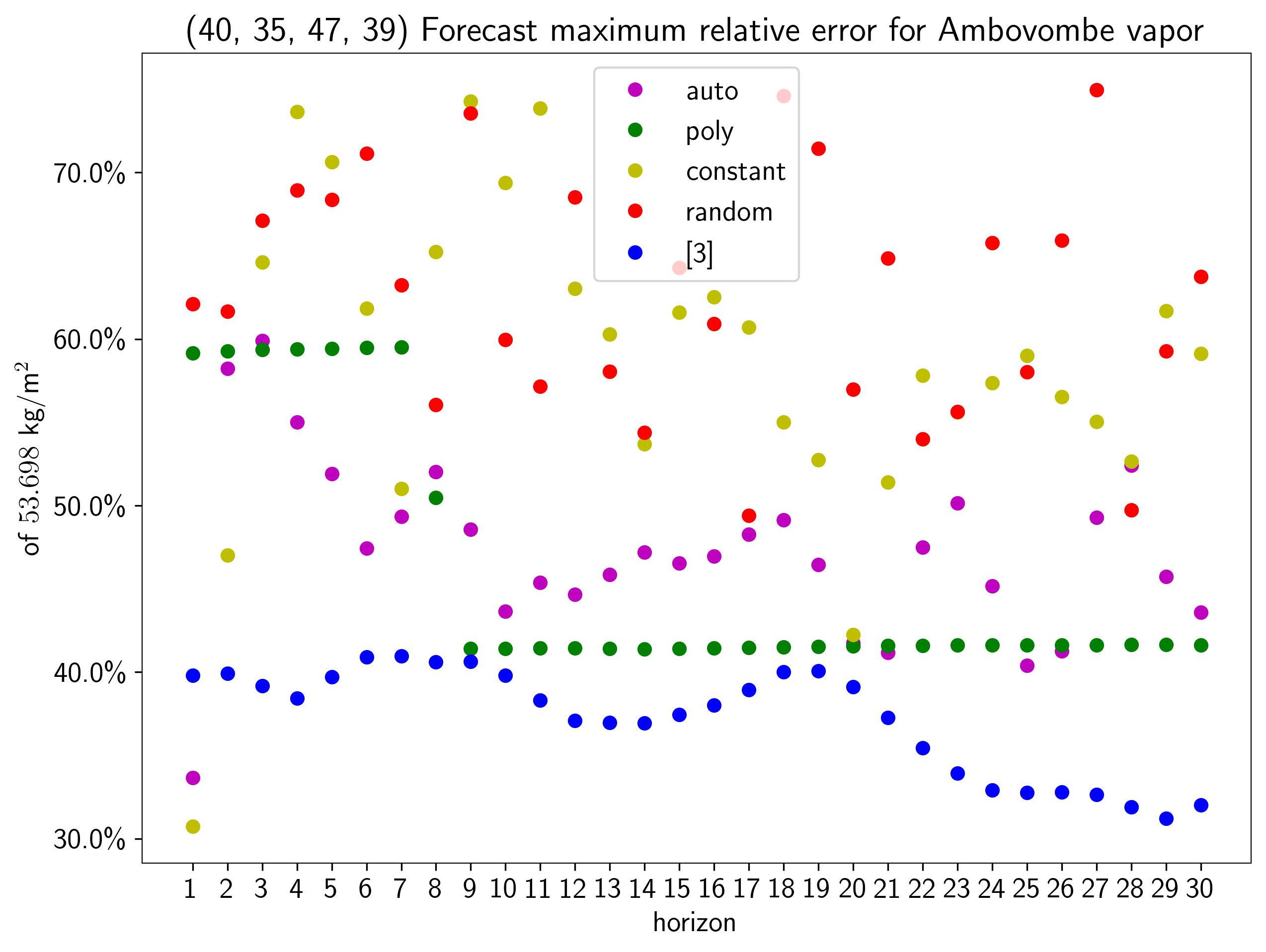}
\includegraphics[width=.5\linewidth]{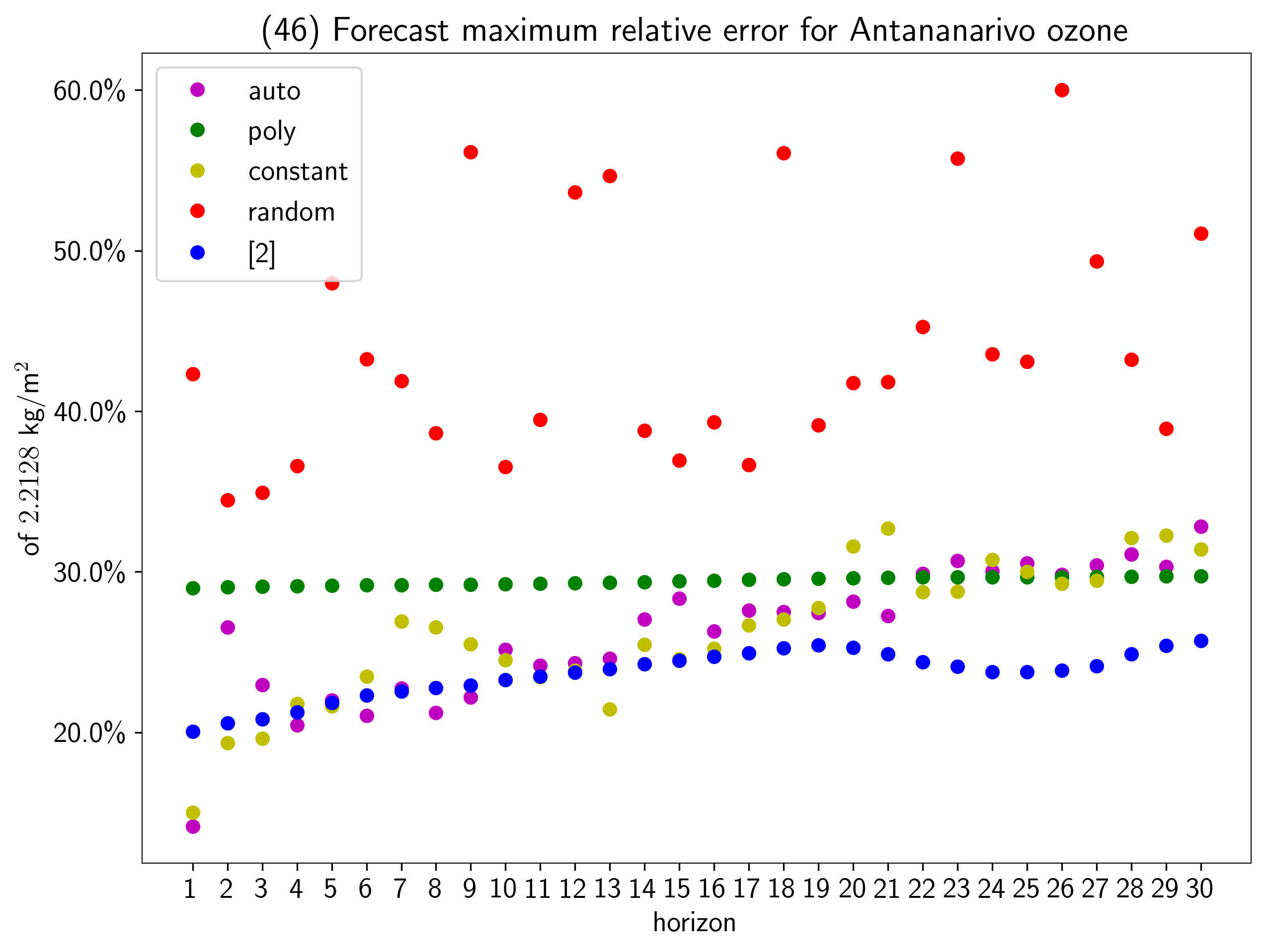}
\includegraphics[width=.5\linewidth]{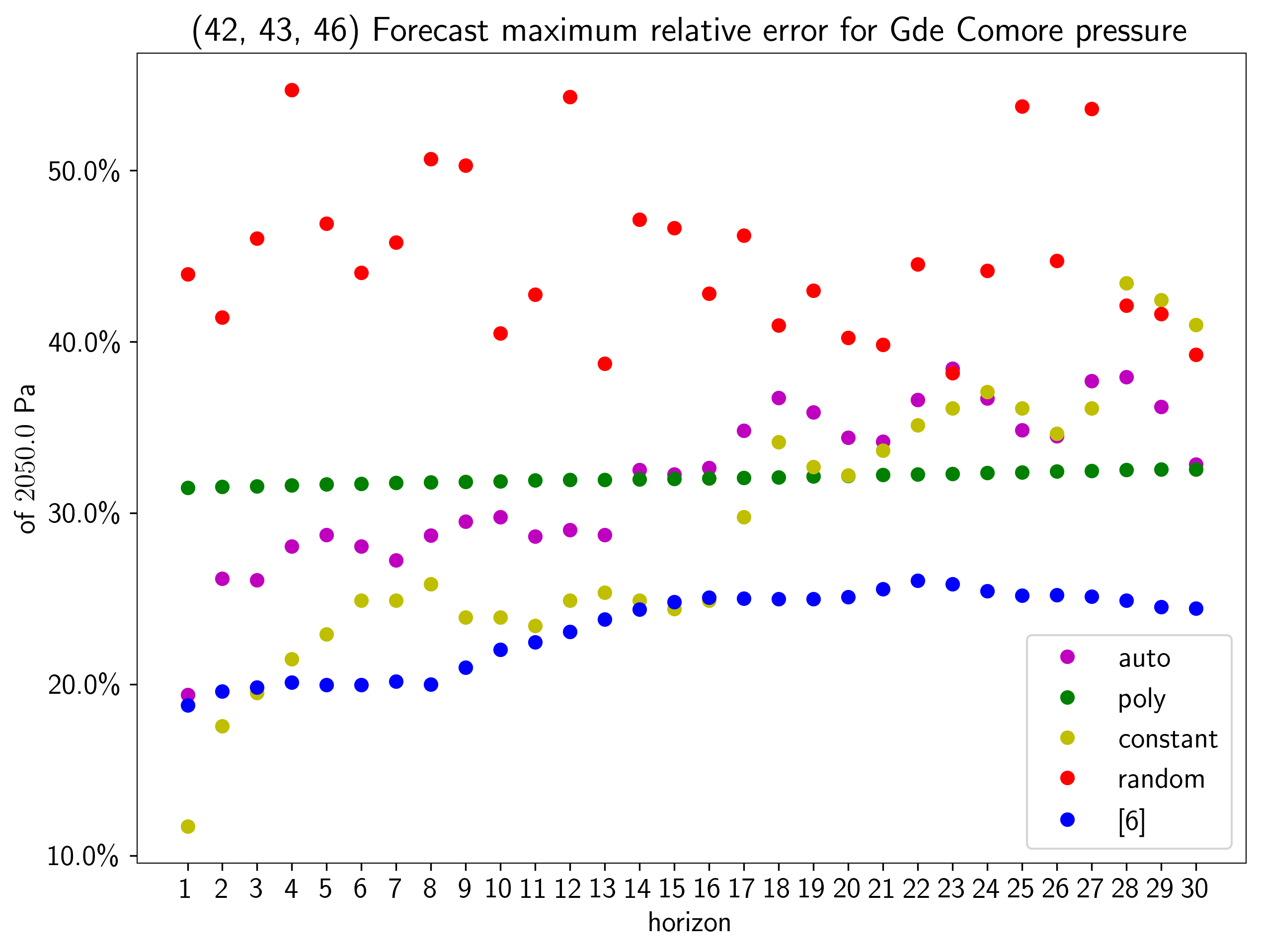}
\section{Comparative plots for two window lengths}
\label{app:compar}
The following plots compare the accuracy of vector forecasting
for window lengths $L^{\text{\cite{maETal:wlen2012}}}$
and $L_{\text{lo}}$. Every label of a legend gives the window size
followed by either ``auto'' for automated grouping using
\textsf{grouping.auto.wcor} or the optimal prefix grouping $[M]$.
For $L_{\text{lo}}$ (resp.\@ $L^{\text{\cite{maETal:wlen2012}}}$)
the accuracy is plotted in blue (resp.\@ green) for automated grouping
and in orange (resp.\@ red) for optimal grouping. The reader may check
Table\nbs\ref{tab:win} to avoid confusion between
$L^{\text{\cite{maETal:wlen2012}}}$ and $L_{\text{lo}}$. Note that
the accuracy obtained with $L_{\text{hi}}$ do not appear on the
following plots as it is significantly worse
than with $L_{\text{lo}}$ and $L^{\text{\cite{maETal:wlen2012}}}$.
\par\noindent
\includegraphics[width=.5\linewidth]{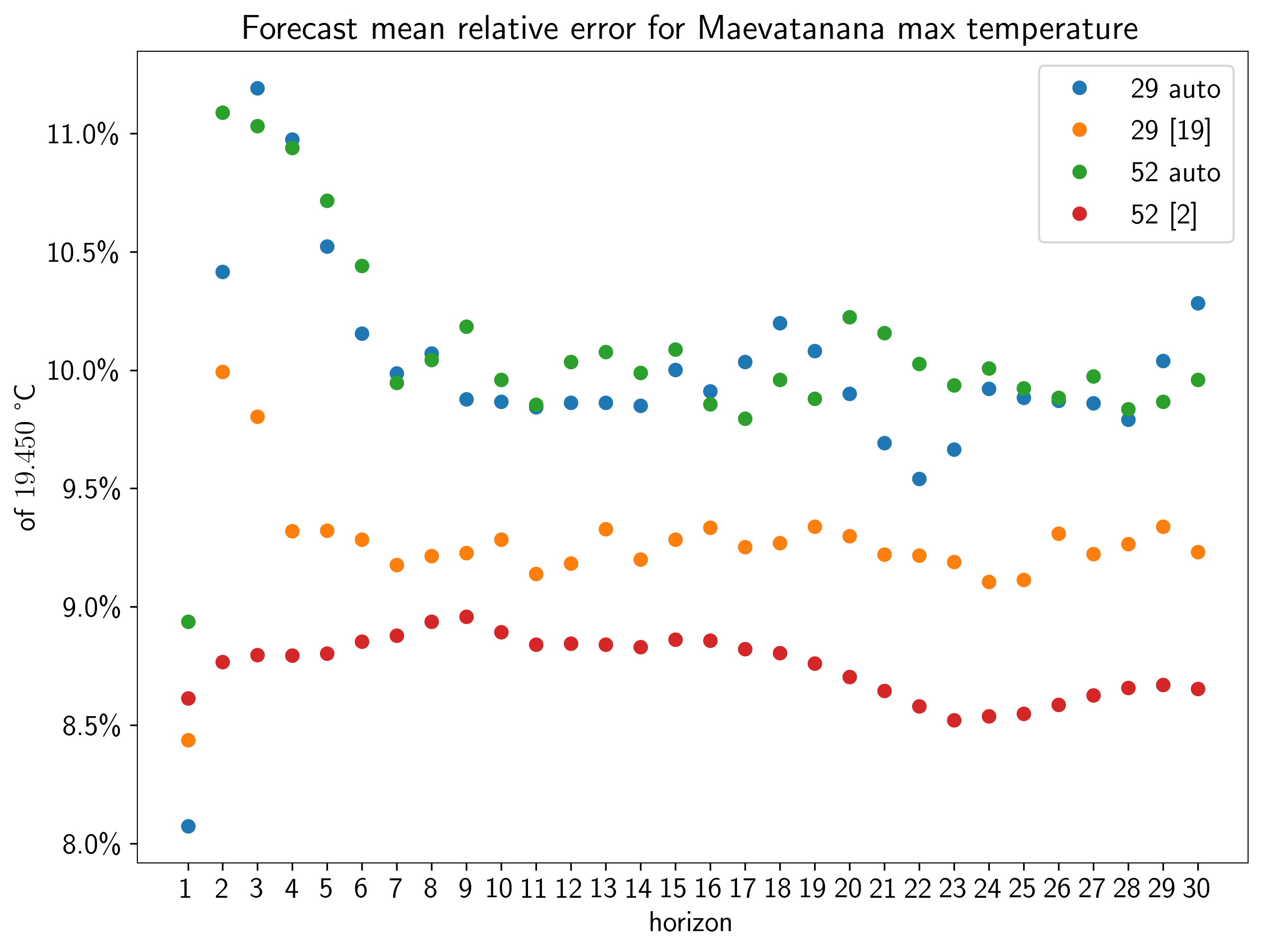}
\includegraphics[width=.5\linewidth]{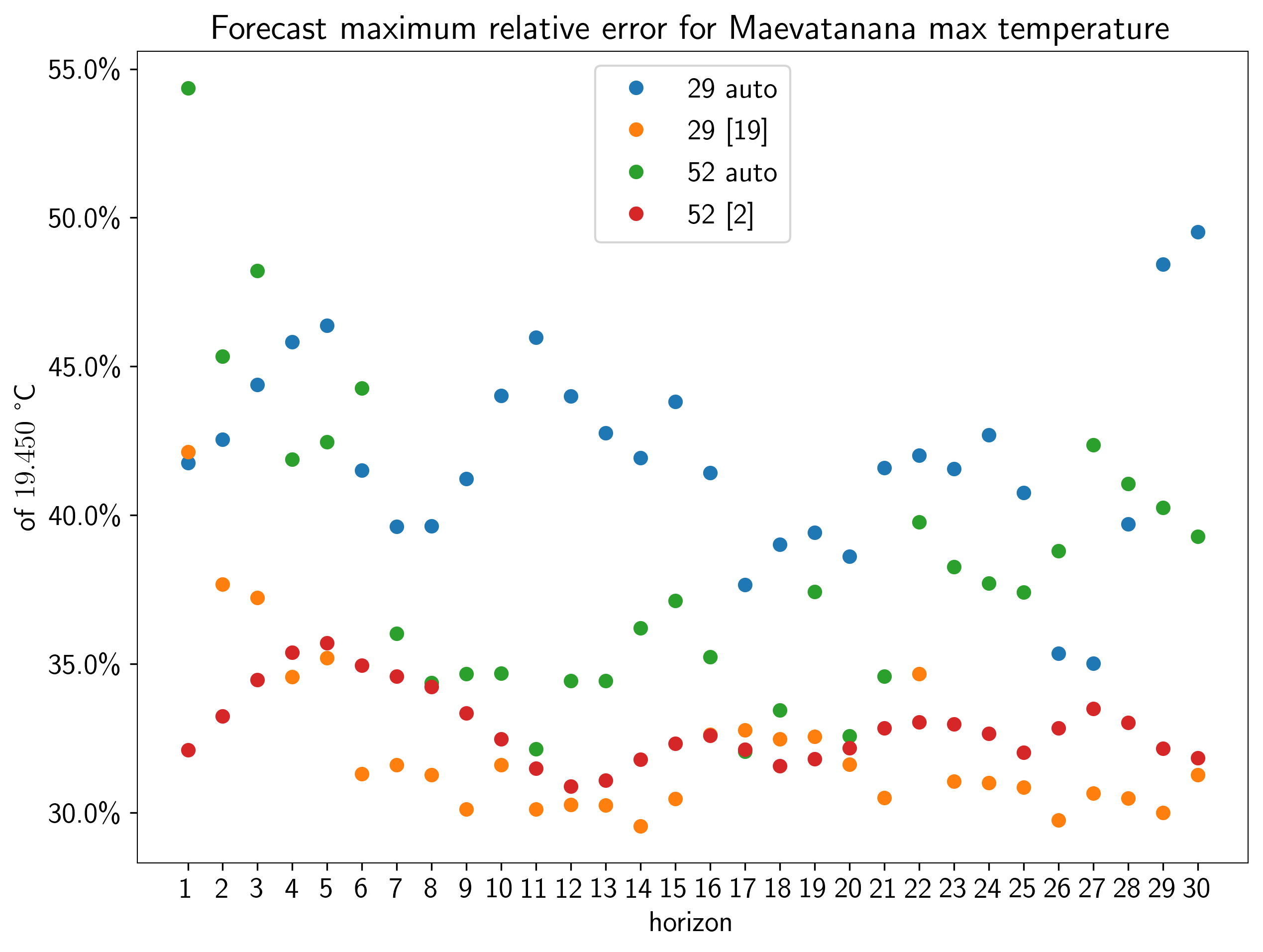}
\includegraphics[width=.5\linewidth]{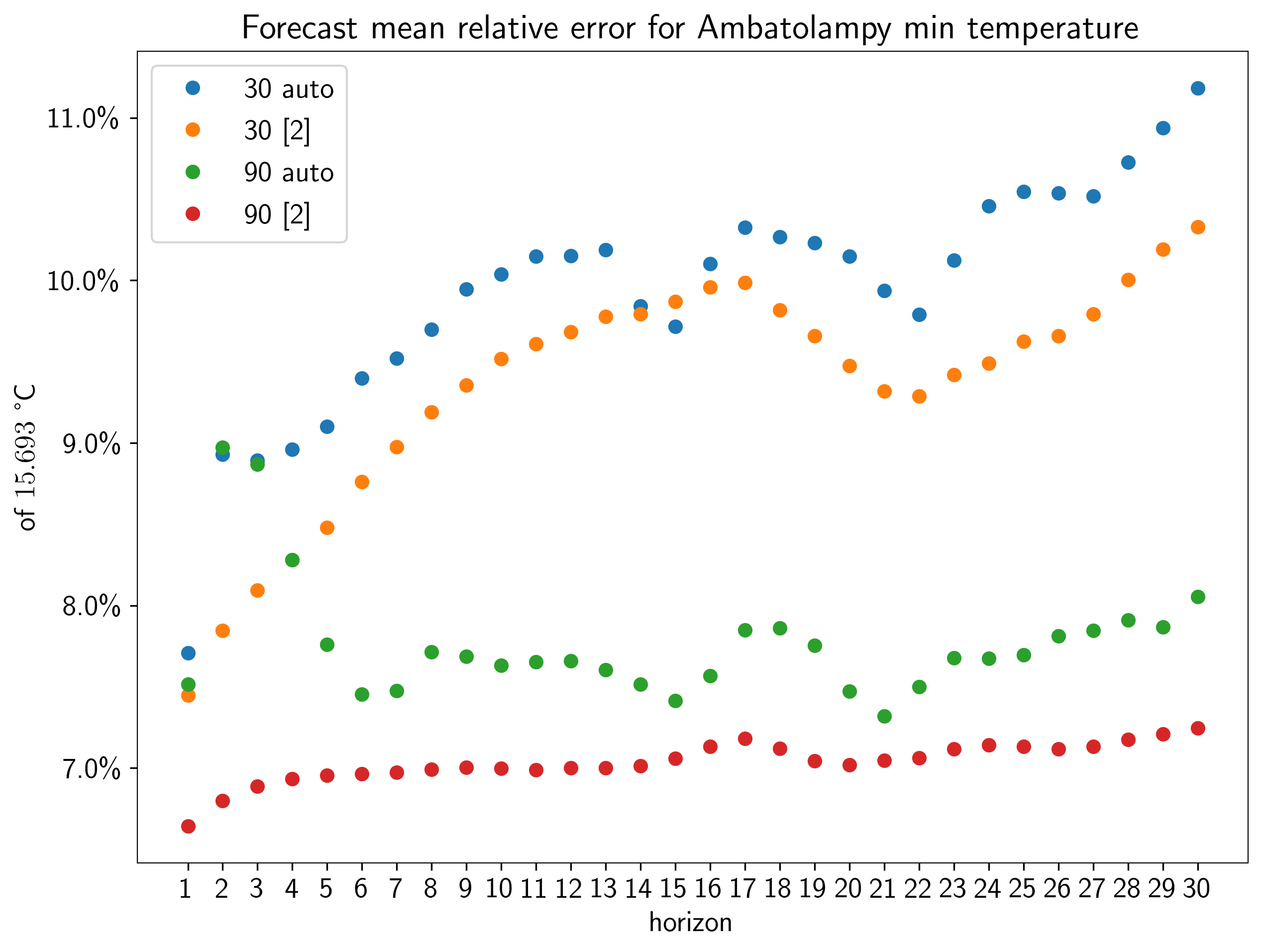}
\includegraphics[width=.5\linewidth]{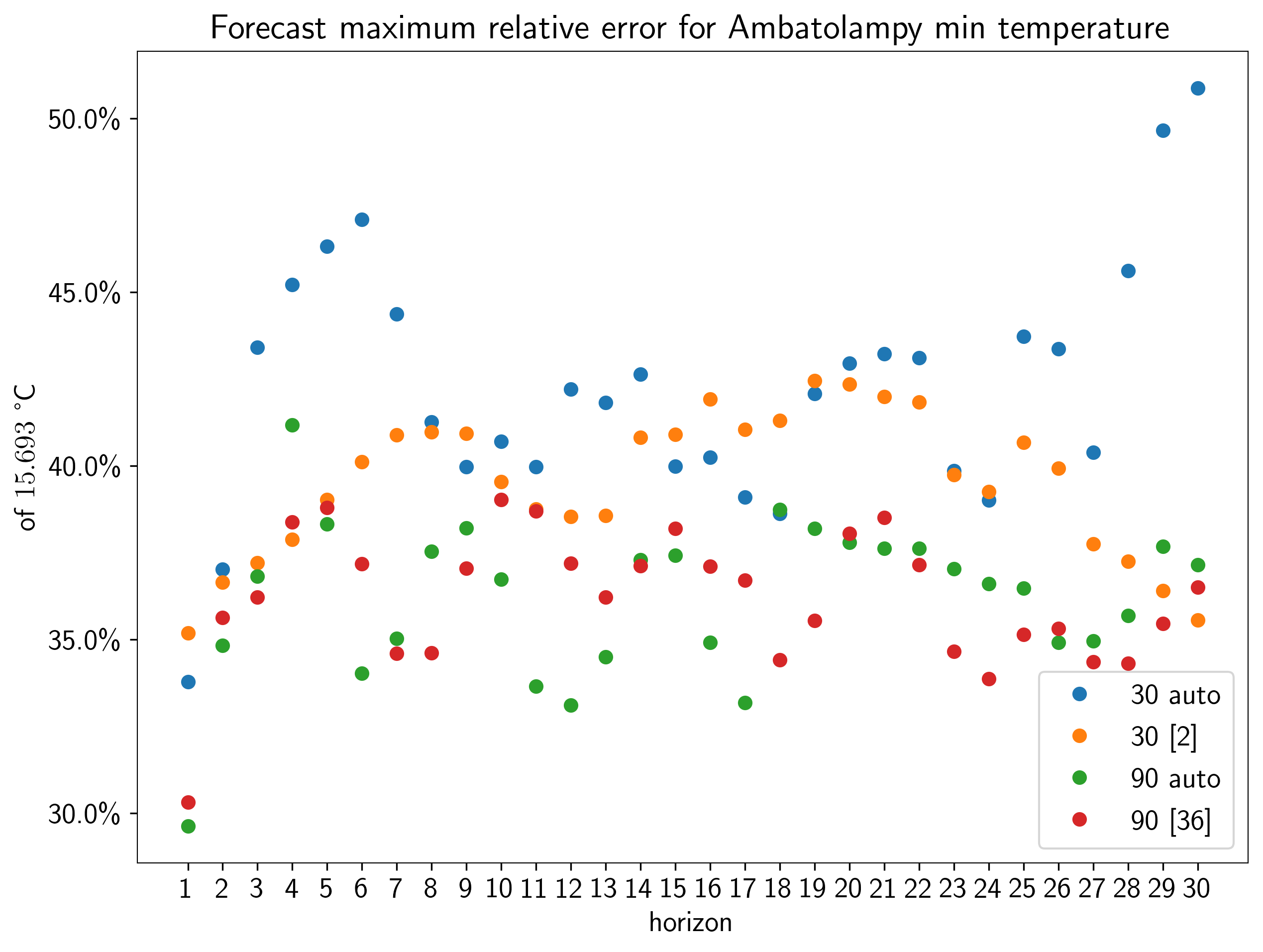}
\includegraphics[width=.5\linewidth]{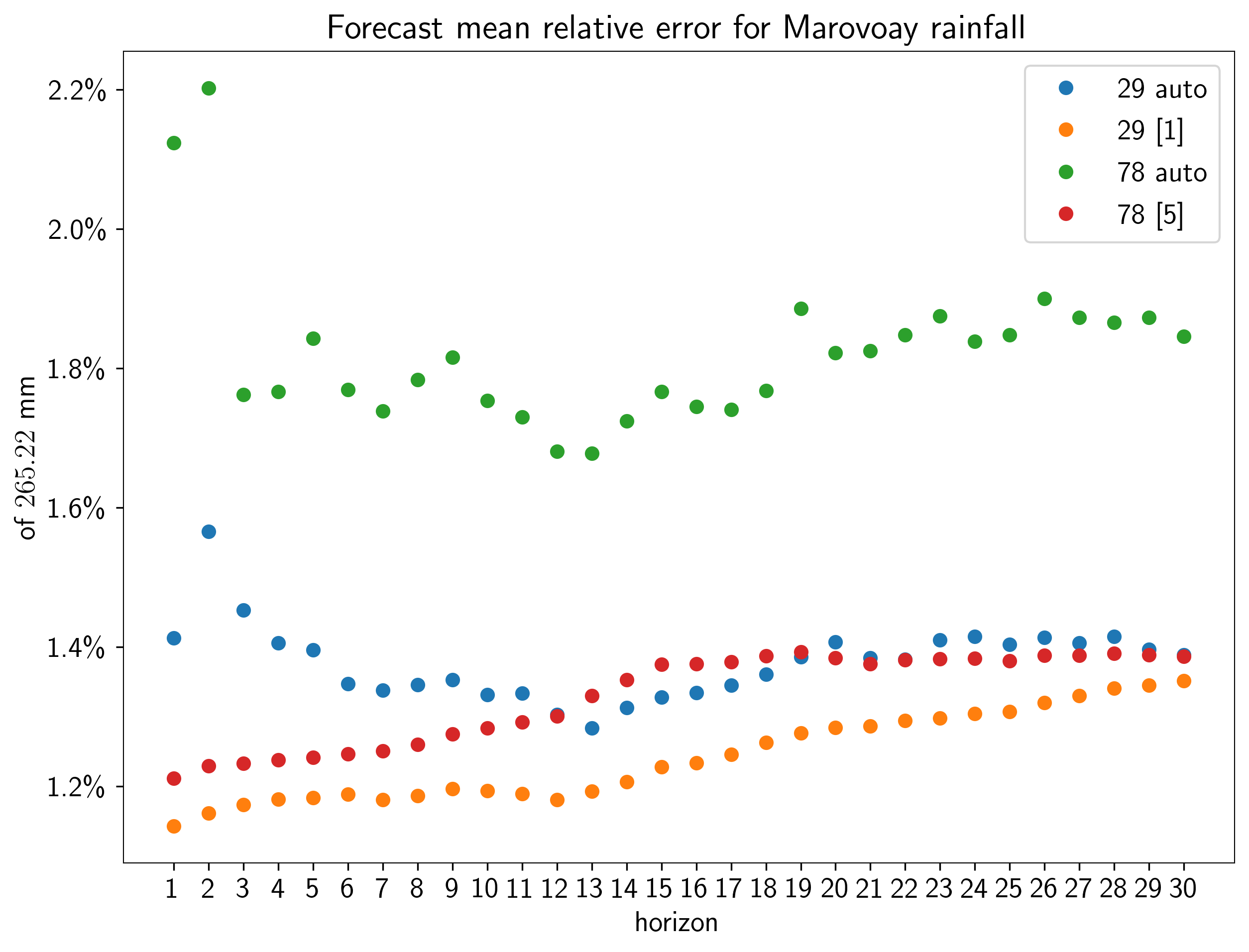}
\includegraphics[width=.5\linewidth]{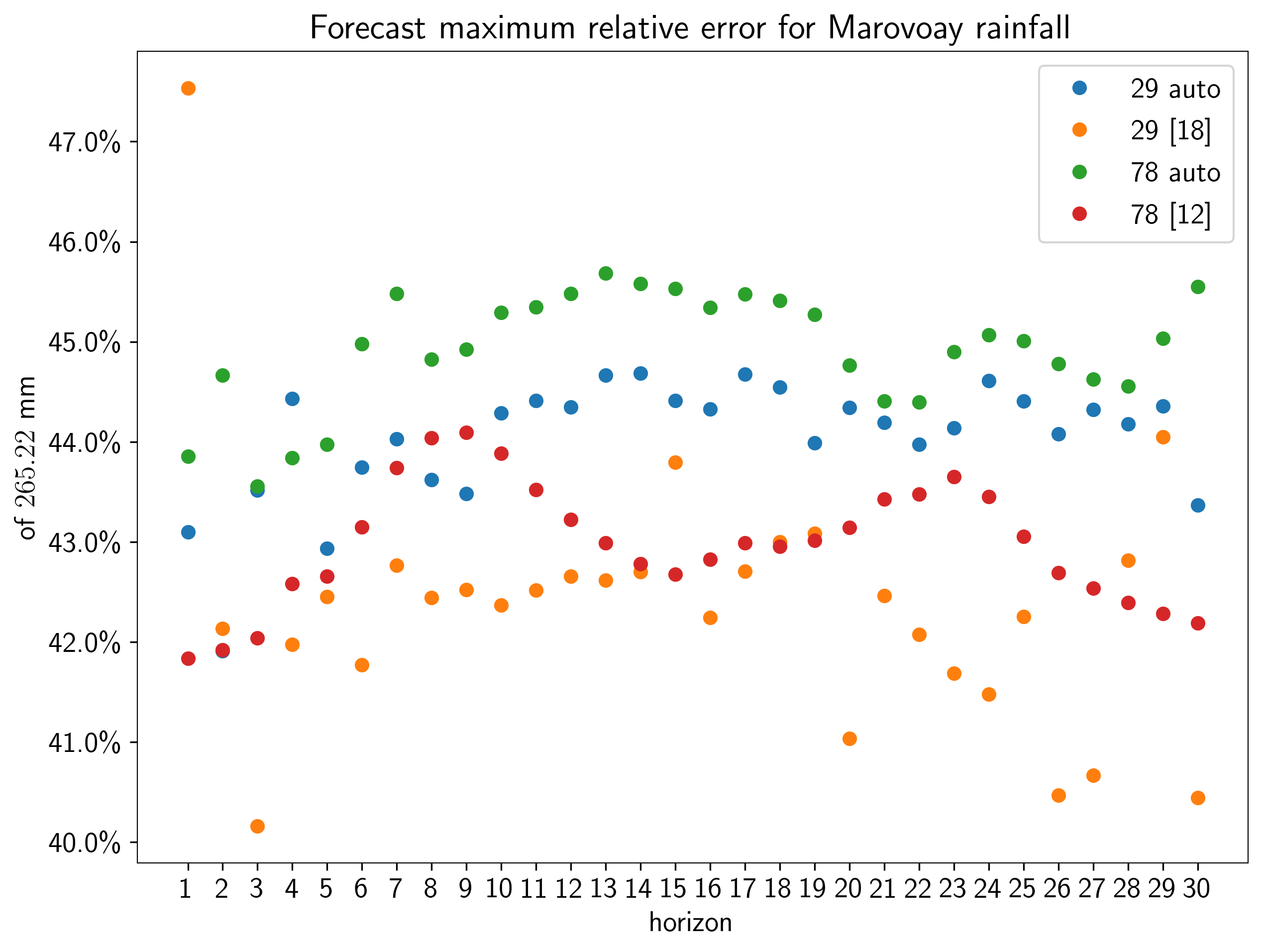}
\includegraphics[width=.5\linewidth]{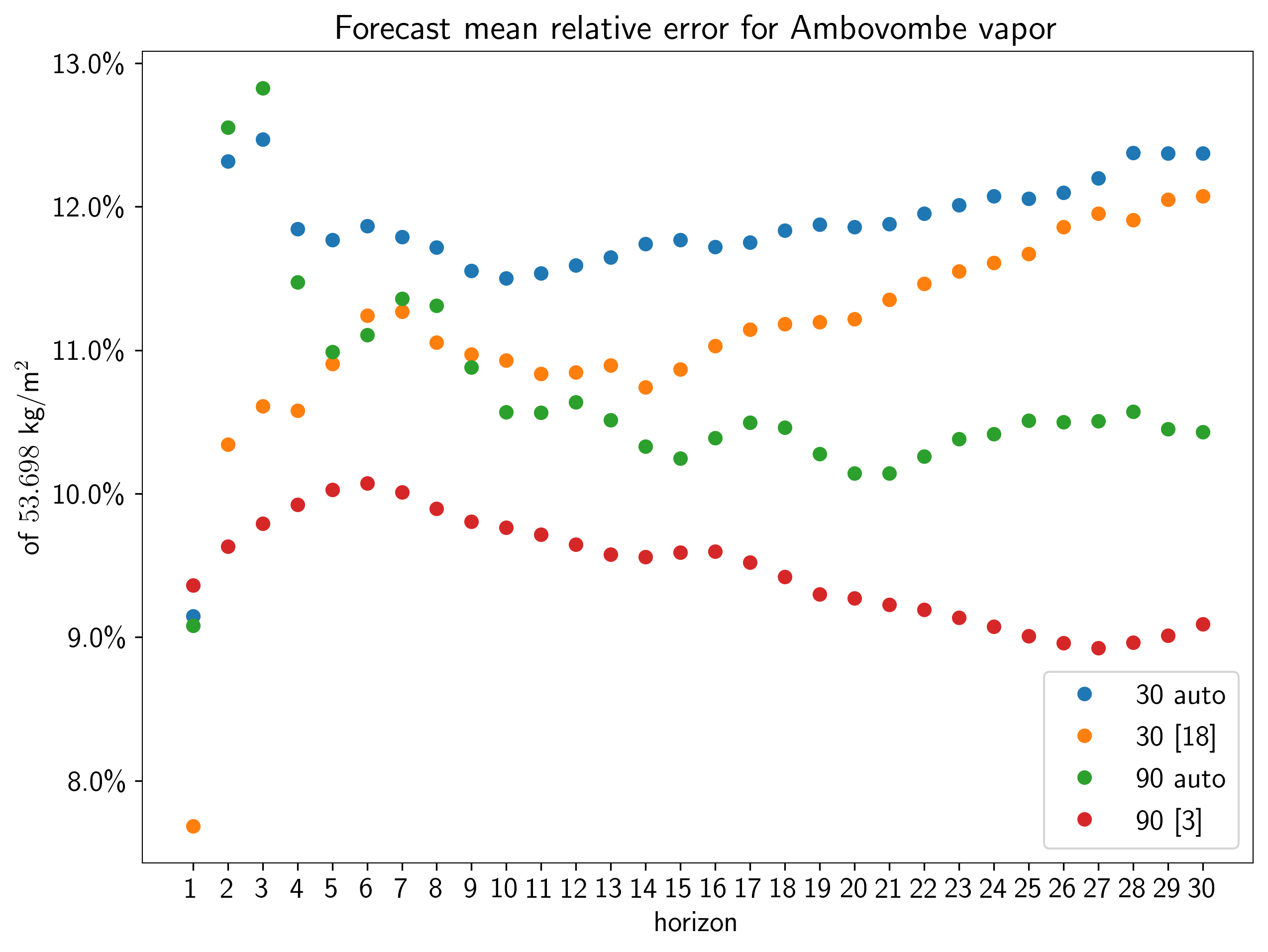}
\includegraphics[width=.5\linewidth]{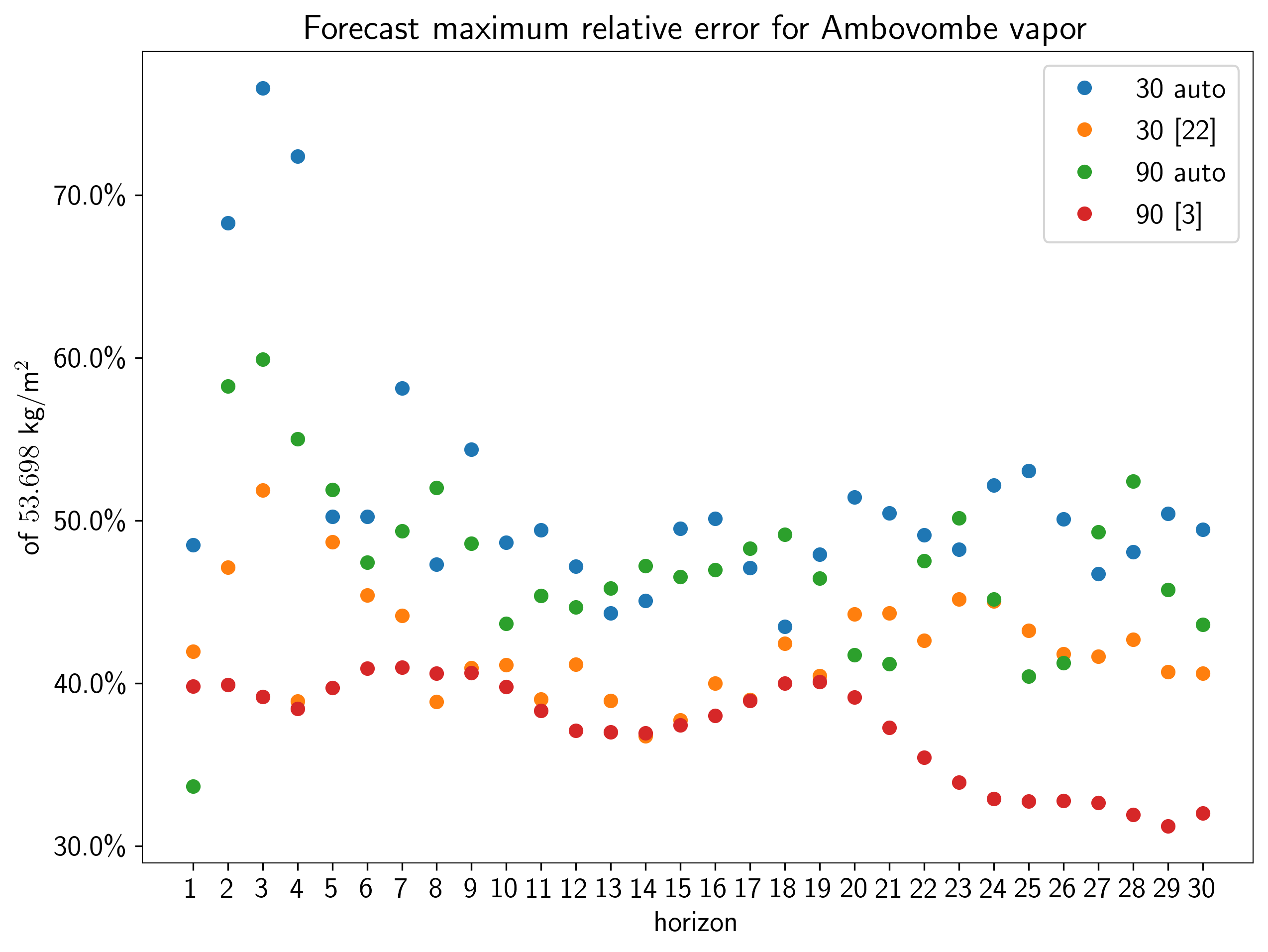}
\includegraphics[width=.5\linewidth]{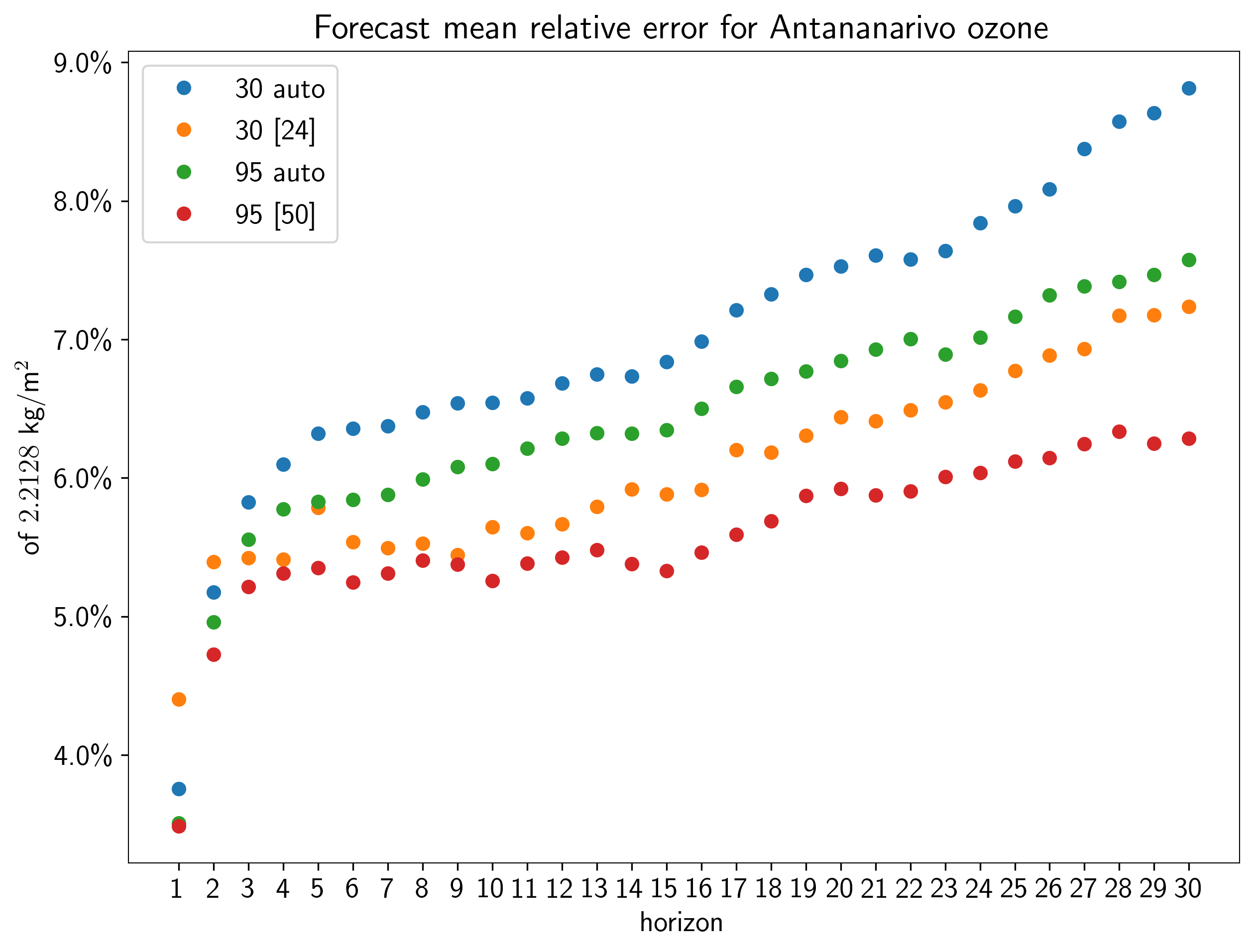}
\includegraphics[width=.5\linewidth]{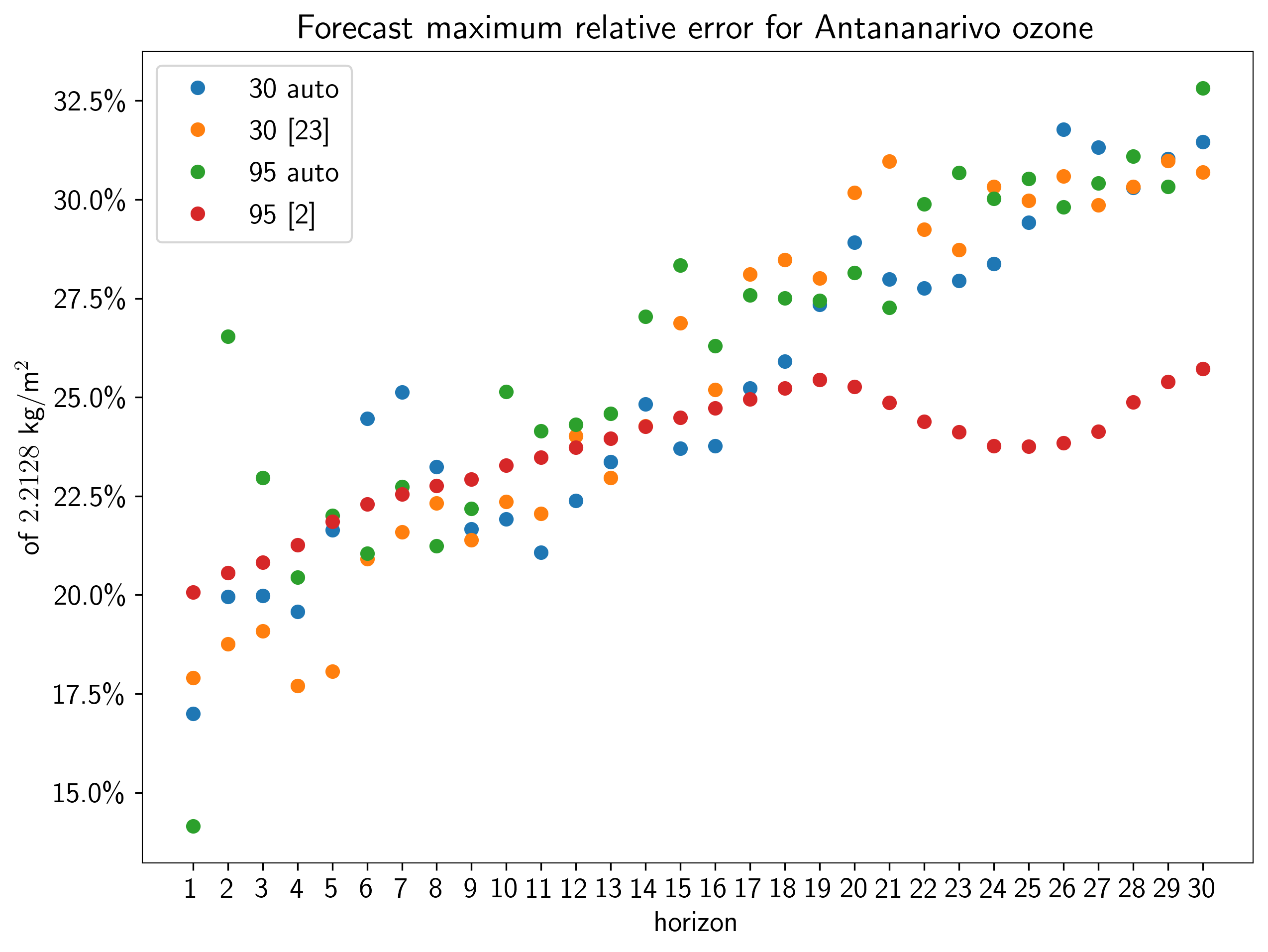}
\includegraphics[width=.5\linewidth]{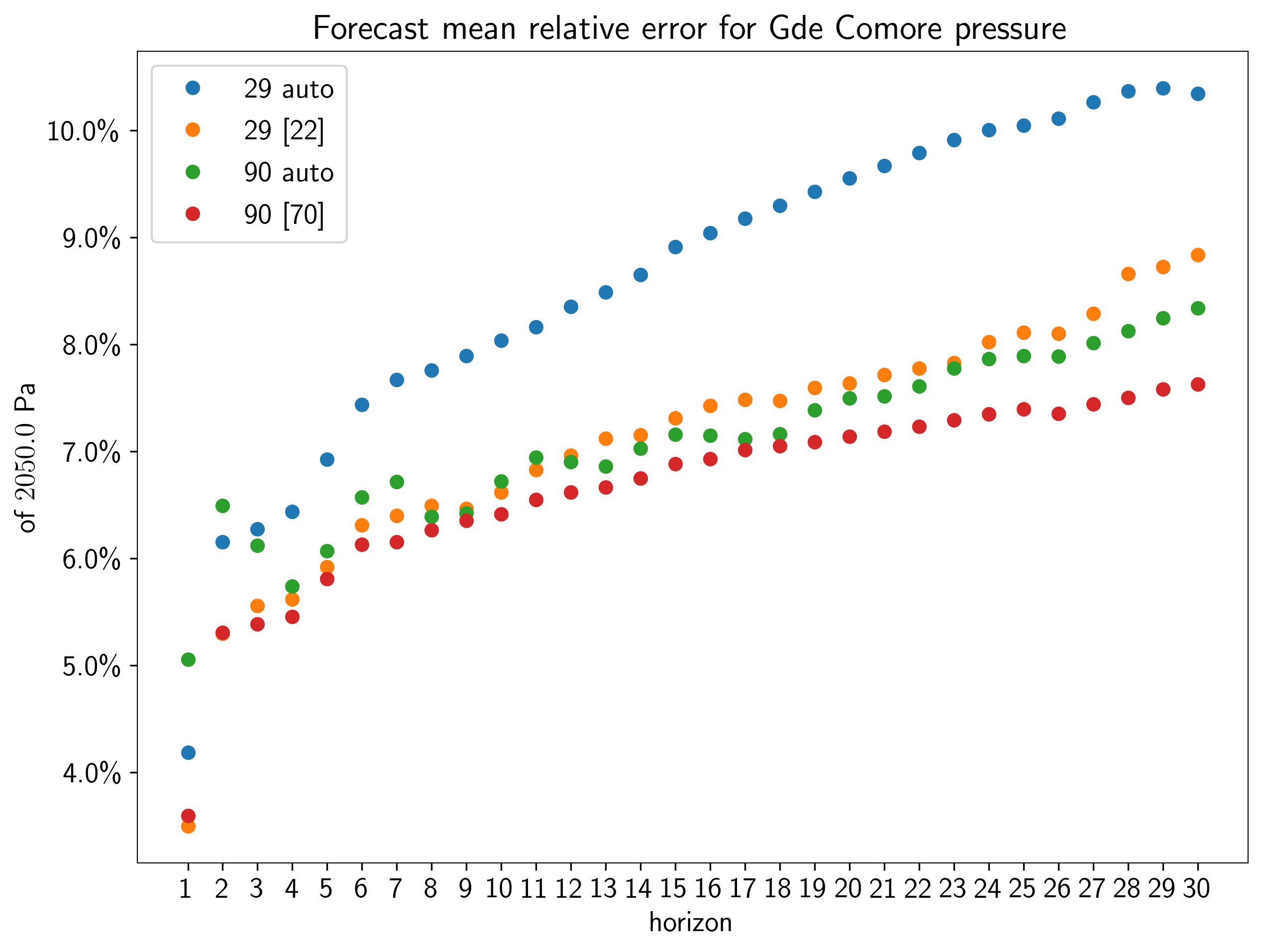}
\includegraphics[width=.5\linewidth]{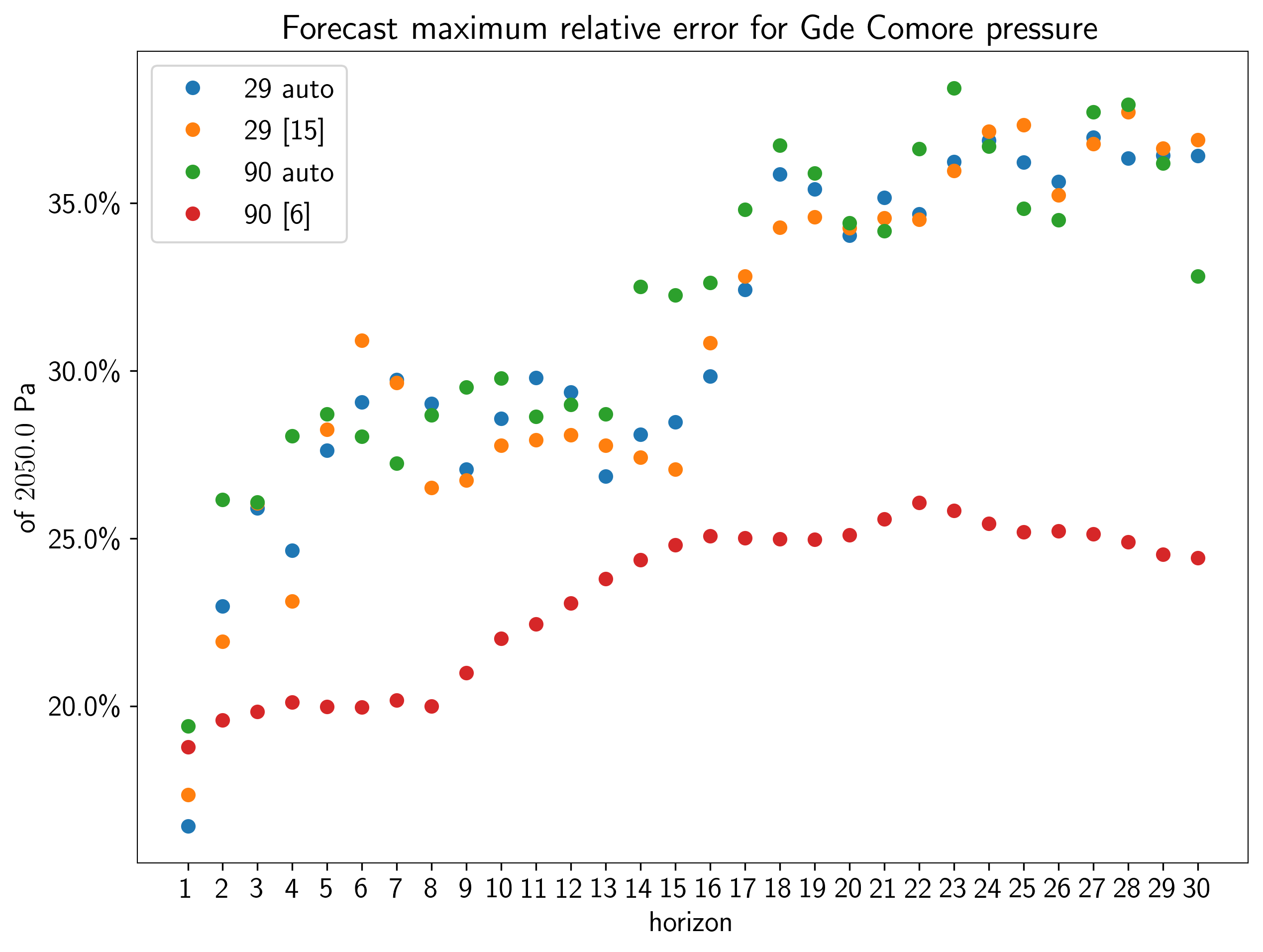}
\end{document}
